\documentclass[useAMS, usenatbib]{mnras}
\usepackage{times}
\usepackage{graphicx}
\usepackage{multirow}
\usepackage{bmpsize}
\usepackage{subcaption}
\usepackage{varioref}
\usepackage{colortbl}
\usepackage{amsmath,environ}
\bibpunct{(}{)}{;}{a}{}{,}

\def\apjl{ApJ}%
\def\apjs{ApJS}%
\def\ao{Appl.~Opt.}%
\def\aap{A\&A}%
\title{Photometric redshifts and clustering of emission line galaxies selected jointly by DES and eBOSS}
\author[S. Jouvel et al.]
{S. Jouvel $^{1}$\thanks{E-mail: sjouvel@ucl.ac.uk}, T. Delubac$^{2}$, J. Comparat$^{3,4}$\thanks{Severo Ochoa IFT Fellow}, 
A. Carnero$^{6,7}$, H. Camacho$^{5,6}$, F. B. Abdalla$^{1,9}$, \and 
J-P Kneib$^{2}$, A. Merson$^{1}$, M. Lima$^{5,6}$, F. Sobreira$^{6,8}$, Luiz da Costa$^{6,7}$, F. Prada$^{3,10,11,12}$, \and 
G. B. Zhu$^{13,14}$, A. Benoit-Levy$^{1}$, A. De La Macora$^{15}$, N.~Kuropatkin$^{8}$, H.~Lin$^{8}$, \and
T. M. C.~Abbott$^{16}$, S.~Allam$^{8}$, M.~Banerji$^{17,18}$, E.~Bertin$^{19,20}$,  D.~Brooks$^{1}$, D.~Capozzi$^{21}$, \and 
M.~Carrasco~Kind$^{22,23}$, J.~Carretero$^{24,25}$, F.~J.~Castander$^{24}$, C.~E.~Cunha$^{26}$, S.~Desai$^{27,28}$, \and 
P.~Doel$^{1}$, T.~F.~Eifler$^{29,30}$, J.~Estrada$^{8}$, A.~Fausti Neto$^{6}$, B.~Flaugher$^{8}$, P.~Fosalba$^{24}$, \and
J.~Frieman$^{8,31}$, E.~Gaztanaga$^{24}$, D.~W.~Gerdes$^{32}$, D.~Gruen$^{33,34}$, R.~A.~Gruendl$^{22,23}$, \and
G.~Gutierrez$^{8}$, K.~Honscheid$^{35,36}$, D.~J.~James$^{16}$, K.~Kuehn$^{37}$, O.~Lahav$^{1}$, T.~S.~Li$^{38}$, \and 
M.~A.~G.~Maia$^{6,39}$, M.~March$^{29}$, J.~L.~Marshall$^{38}$, R.~Miquel$^{25,40}$, W.J.~Percival$^{21}$,\and
A.~A.~Plazas$^{30}$, K.~Reil$^{41}$, A.~K.~Romer$^{42}$, A.~Roodman$^{26,42}$, E.~S.~Rykoff$^{26,42}$, M.~Sako$^{29}$,\and
E.~Sanchez$^{43}$, B.~Santiago$^{6,44}$, V.~Scarpine$^{8}$, I.~Sevilla-Noarbe$^{22,43}$, M.~Soares-Santos$^{8}$,\and
E.~Suchyta$^{35,45}$,
G.~Tarle$^{32}$,
J.~Thaler$^{46}$,
D.~Thomas$^{21}$,
A. Walker$^{16}$,
Y.~Zhang$^{32}$,
J.~Brownstein$^{47}$
\vspace{0.4cm}\\
Affiliations are listed at the end of the paper
}
\begin{document}
\maketitle
\begin{abstract} 
We present the results of the first test plates of the extended Baryon Oscillation 
Spectroscopic Survey. This paper focuses on the emission line galaxies (ELG) 
population targetted from the Dark Energy Survey (DES) photometry. 
We analyse the success rate, efficiency, redshift distribution, 
and clustering properties of the targets. 
From the 9000 spectroscopic redshifts targetted, 4600 have been selected from 
the DES photometry. The total success rate  
for redshifts between 0.6 and 1.2 is 71\% and 68\% respectively for a bright 
and faint, on average more distant, samples including redshifts measured from a single strong emission line.
We find a mean redshift of 0.8 and 0.87, with 15 and 13\% of unknown 
redshifts respectively for the bright and faint samples. 
In the redshift range $0.6<z<1.2$, for the most secure spectroscopic redshifts, 
the mean redshift for the bright and faint sample is 0.85 
and 0.9 respectively. Star contamination is lower than 2\%. We measure a galaxy bias 
averaged on scales of $1$ and $10$~Mpc$/h$ of $1.72 \pm 0.1$ for the bright sample 
and of $1.78 \pm 0.12$ for the faint sample. The error on the galaxy bias have been obtained propagating the 
errors in the correlation function to the fitted parameters. This redshift evolution for the galaxy 
bias is in agreement with theoretical expectations for a galaxy population with 
$M_{B}-5\log h < -21.0$. We note that biasing is derived from the galaxy clustering 
relative to a model for the mass fluctuations. 
We investigate the quality of the DES photometric redshifts and 
find that the outlier fraction can be reduced using a comparison between template fitting
and neural network, or using a random forest algorithm.
\end{abstract}

\begin{keywords}
Cosmology: observations -- Surveys -- target selection -- DES -- redshift
\end{keywords}

\section{Introduction}
\label{sec:intro}
With the development of new technologies, instruments improving in their performance, we are planning wider and
deeper surveys with a volume of an order of four magnitudes higher than 20 years ago.  
Our surveys fall into two categories: spectroscopic redshift surveys for galaxy clustering 
and photometric redshift surveys for gravitational lensing, clusters, and clustering. 
Photometric and spectroscopic surveys mutually need each other in the sense that spectroscopic
surveys need photometry to estimate galaxy properties such as colors, stellar masses, approximate redshifts, 
but also to efficiently select and understand its targets such as biasing.
Photometric surveys need spectroscopic data to quantify the accuracy of photometric redshift. 
Both aspects are subjects of this paper.

We first present the characteristics of the emission line galaxy samples observed with the  
extended Baryon Oscillation Spectroscopic 
Survey \footnote{https://www.sdss3.org/future/eboss.php}, (hereafter eBOSS). These data are part of the
eBOSS ELG target selection definition effort, undergone in October 2014. 
We designed different target selections based on Sloan Digital Sky Survey (SDSS) \citep{Ahn14}, 
the South Galactic Cap u-band Sky Survey (SCUSS) \citep{Jia14}, and the Dark Energy Survey\footnote{http://www.darkenergysurvey.org/},
hereafter DES.\\
eBOSS is a spectroscopic survey using the BOSS spectrograph \citep{Smee13} at the Apache Point Observatory. 
It will cover 7500deg$^2$ in a six-year period starting Fall 2014. 
eBOSS aims at measuring the baryon accoustic oscillation feature at redshift higher than 0.6 
extending the first measurement from the SDSS at lower redshift \citep{Percival09}. eBOSS will
use a mixture of targets to have a measurement at z$\sim$1 using high redshift emission line galaxies (ELG), 
quasars between redshift one and two \citep{Leistedt14}, and Lyman-alpha absorption quasars at redshift higher than two \citep{FontRibera14}. 
In this paper, we present the results of two high redshift ELGs target selections and test plates spectroscopic redshift
distributions using the DES photometry. We also study the bias for DES-based target selections using the DES Science Verification data.
Two companion papers study other photometric selections \citep{Delubac15} in preparation,
and luminosity functions \citep{Comparat15} in preparation. \citet{Comparat15} in preparation gives full details about this test
observations detailing the pipeline of redshift reduction.  

Several papers study the survey design of efficient target selection of ELGs
such as \citet{Comparat13,Adelberger04} as well as observations such as DEEP2 \citep{Newman13},
VIPERS \citep{Garilli14} and Wigglez \citep{Parkinson12}. DESI \citep{Schlegel11} and eBOSS are
a new generation of surveys using the latest technologies in the field of spectroscopy. 
Using these new instruments allows to cover 
large sky areas selecting higher redshift and fainter targets than preceding surveys, 
increasing the statistical confidence in the measurement of cosmological parameters.\\ 

On the photometric side, DES is an ongoing photometric ground-based galaxy survey which started 
in autumn 2013. DES uses the brand new 2.2 deg$^2$ DECam instrument\citep{Flaugher15} mounted on the 4m Victor M. Blanco Telescope located 
at the Cerro Tololo Inter-American Observatory (CTIO) in Chile. It will cover 5000deg$^2$ after completion
in five optical broad bands observing the southern sky. DES will use cosmic shear, cluster counts, large scale structure measurements
and supernovae to reach very competitive measurement of the Universe growth rate and dark energy. \\ 
One of the main limitations in ongoing and upcoming dark energy photometric surveys are the access
to the radial dimension, the redshift. Surveys such as DES, LSST, Euclid, have between 5 to 8 photometric broad bands.
Broad band photometric redshifts ($z_{ph}$) have an accuracy limited by 
the filter resolution \citep{Jouvel11}. In the optimal case of a space-based survey with 
16 filters covering from UV to infrared, photometric redshifts reach a precision of 
0.03 \citep{Jouvel14CLASH}, in the case of crowded fields, with cluster insterstellar light.  
Photometric redshifts precision is the unclipped rms of the redshift measurement error, the difference
between the measured and true redshift. 
Depending on the survey configuration such as ground-based, space-based,
number of broadband filters, pixel size and exposure time, the photometric redshifts accuracy will go from
a bit better than 0.03 and degrade. DES reaches a precision of 0.08 \citep{Sanchez14} but this also depends on the color, magnitude, size, and 
redshift of the galaxy population considered. 
If photometric redshifts are used to measure galaxy clustering, then the resulting estimate of the dark matter
power spectrum becomes biased at small scales because the density fluctuation traced by galaxies appear 
to be smoothed by the photometric redshift error.
The bluest galaxies show a spectrum with higher degeneracies in color space allowing
some high redshift galaxies to be misplaced at low redshift, and conversely.
Dark Energy Equation of State constraints, which rely on photometric redshift information 
(like weak-lensing and cluster mass function estimates) can be severely affected by (unaccounted)
outliers \citep{Bernstein09}.
In this paper we use the eBOSS test plates redshifts to investigate possible ways to reduce the outlier fraction. 

In section \ref{sec:test_plates}, we present the definition, efficiency and spectrocopic characteristics of our DES 
photometrically selected sample, based on the DES Science Verification data \citep{Melchior14,Banerji15,Sanchez14}.
Using the eBOSS spectroscopic redshifts of the test observation we look at the photometric 
redshift accuracy of the DES survey and outlier removal in section \ref{sec:photoz}. 
In section \ref{sec:final_targets}, we show the results of our systematics studies on the year one DES data which is an order
of magnitude larger in area than the test plates presented in this paper.
In section \ref{sec:clustering} we focus on the clustering properties of the spectroscopic samples and the impact of 
photometric redshift precision in the clustering from the point of view of DES. Assuming that the
DES photometric redshift distribution are well calibrated, we study the impact of possible biases
coming from the bins boundaries, due to photometric redshits. Finally we present the conclusions 
on the characterization, suggestions for the improvement of the photometric DES target selection, 
and the outlier removal techniques. 

\section{DES and eBOSS data}
\label{sec:test_plates}
%%%%%%%%%%%%%%%%%%%%%%%%%%%%%%%%%%%%%%%%%%%%%%%%%%%%%%%%%%%%%%%%%%%%%%%%%%%%%%%%%%%%%%%%%%%%%%%%%%
\begin{figure*}
\includegraphics[width=0.7\textwidth,height=0.4\textwidth]{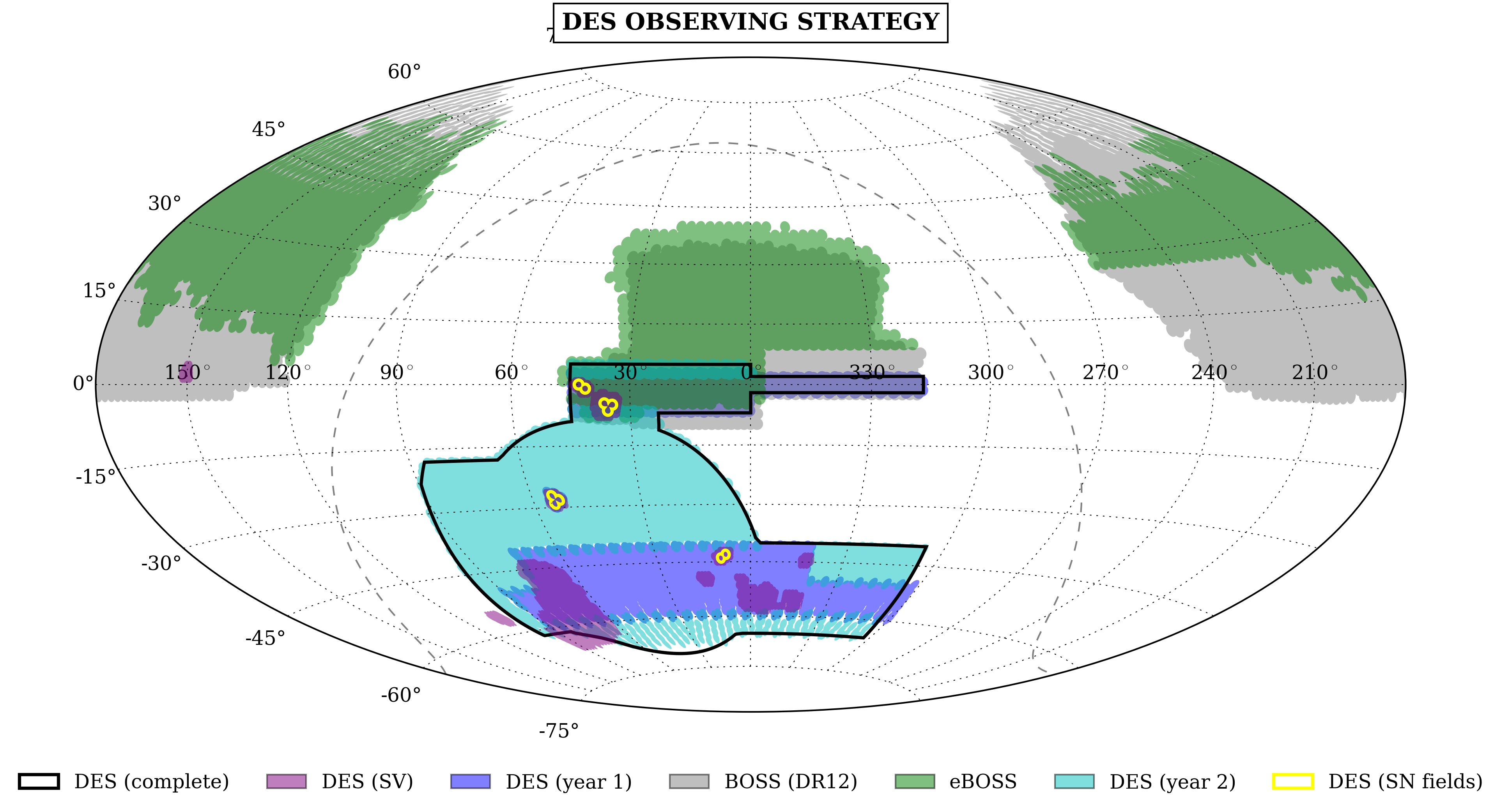}
\caption{Footprints of DES, BOSS, and eBOSS. Coordinates are RA and Dec in deg.} %\textcolor{blue}{Maybe only plot the overlap regions ?}}
\label{fig:footprint}
\end{figure*}
%%%%%%%%%%%%%%%%%%%%%%%%%%%%%%%%%%%%%%%%%%%%%%%%%%%%%%%%%%%%%%%%%%%%%%%%%%%%%%%%%%%%%%%%%%%%%%%%%%

The eBOSS fields selected for the ELG target selection campaign were chosen to overlap with CFHTLS-W1.
CFHTLS has four wide and four deep 1deg$^2$ fields in u*,g',r',i',z' bands. The W1 field is the
biggest of the wide fields with 19deg$^2$ and 80\% completness depth of i'$<$24.5. 
\citet{Coupon09} computed the CFHTLS photometric redshift accurate to 3-4\% up to i'$<$22.5 calibrated with VVDS \citep{LeFevre05},
DEEP2 \citep{Newman13} and zCOSMOS \citep{Lilly07} spectroscopic surveys. 
These data are publicly available.  
This field has also been imaged by the SDSS survey \footnote{http://www.sdss.org/data}. The photometry
from SDSS is a 7500 deg$^2$ with 95\% completness depth of  $u, g, r, i, z = 22.0, 22.2, 22.2, 21.3, 20.5$
\citep{Abazajian09}. 
 
eBOSS and DES have an overlap of 500 deg$^2$ over Stripe82 which will yield a minimum number of 60000 spectra.
Since DES photometry is deeper than SDSS, we optimise the target
selection to reach fainter targets at higher redshift and lower contamination from low redshift galaxies. 
\citet{Abdalla08nn} showed that a neural network can pick up strong emission lines from the broad-band photometry      
of a galaxy using DEEP2 and SDSS data. Further optimisation has been done applying this method
to a DESI-like survey in \citet{Jouvel14}. \citet{Jouvel14} shows that one reaches a higher success rate
using neural networks target selections than boxes in color space.   
A related work used a Fisher discriminant method to investigate improvements in the eBOSS target selection \citep{Raichoor15}.
We note that success rate here is the percentage of objects for which we have a very secure galaxy redshift. 
The completness of the target selection in terms of galaxy population would be the subject of an other paper. 
%Spectrocopic redshift will be used to calibrate the redshift distribution
%for cosmic shear measurement. 
%The benefit from overlapping spectroscopic and photometric survey has been studied in diverse
%papers \citep{Kirk13} showing a large range of results depending on the assumptions going 
%in the calculations. Studying this overlapping region will certainly be
%interesting.
%%%%%%%%%%%%%%%%%%%%%%%%%%%%%%%
\begin{figure*}
\includegraphics[width=1\textwidth,height=0.3\textwidth]{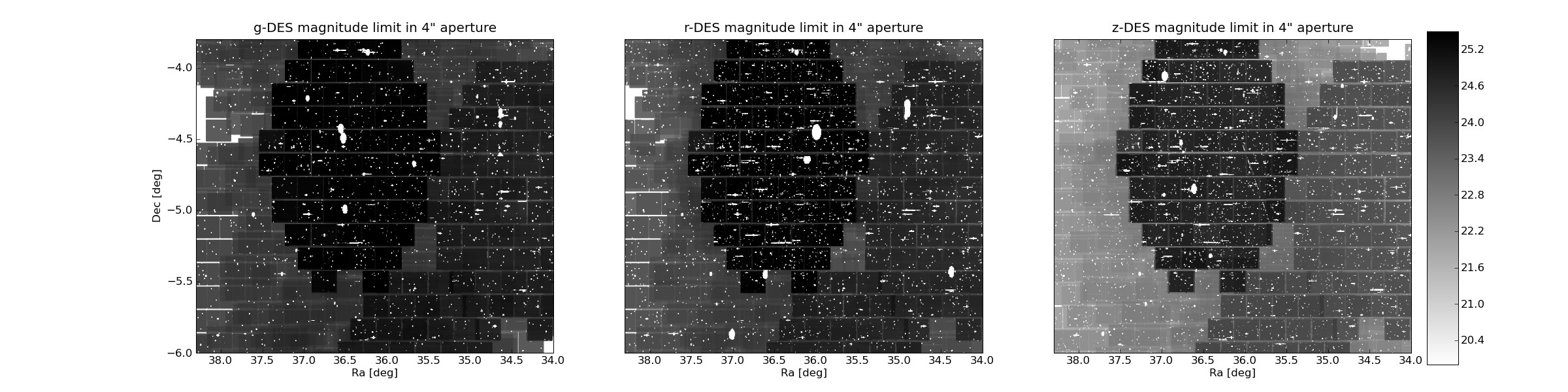}
\caption{Depth of the g,r,z-band of the DES Science Verification data in the 9.2 deg$^2$ of the eBOSS test plates. 
The depth has been computed with the Mangle software and corresponds to a 10$\sigma$ magnitude in an aperture of 2 arcsec.
Find more details in section \ref{sec:clustering}.}
\label{fig:grz_depth_eboss}
\end{figure*}
%%%%%%%%%%%%%%%%%%%%%%%%%%%%%%%%%%%%%%%%%%%%%%%%%%%%%%%%%%%%%%%%%%%%%%%%%%%%%%%%%%%%%%%%%%%%%%%%%%

\subsection{DES Science Verification data}
The first phase of the DES survey consisted of various tests and improvements in the data acquisition, 
instrument calibration and data processing, which resulted in a first well defined source catalog, the
Science verification data, hereafter SVA1. 
Scientific results from \citet{Bonnett15,Rozo15,Crocce15,Banerji15,Sanchez14,Melchior14} and others show the very
good quality of the SVA1 data, going at the nominal depth of the DES survey.
In Figure \ref{fig:footprint} we show the footprint of
the DES, BOSS and eBOSS surveys along with the DES year one data, and the SVA1 data. 

To define the target selection of the eBOSS ELG sample, explained in the next subsection, we used an area of 9.2 deg$^2$ from the SVA1 data. 
Figure \ref{fig:grz_depth_eboss} shows the depth over the 9.2 deg$^2$ SVA1 footprint in g,r,z used for the eBOSS observations. 
Limiting magnitudes are defined by the flux in a 4 arcseconds aperture above 10 $\sigma$, computed from the DES images using the Mangle software \citep{Mangle12}. 
Details about the resulting spectroscopic sample will be presented in subsection \ref{sec:final_targets}.

In section \ref{sec:photoz} we explore the photometric redshifts ($z_{ph}$) for our sample. 
We use the ANNz2 code \citep{Sadeh15}, which is a new version of ANNz \citep{Collister04}.
We also show photometric redshifts from the template fitting code Le Phare \citep{Ilbert06,Ilbert09}. 
We computed zeropoints corrections with the ELG samples and applied them to the Le Phare results. 
Except from this, the Le Phare parameters used here are the same than the one used in \citet{Sanchez14}. 
The random forest code TPZ\footnote{ http://lcdm.astro.illinois.edu/code/mlz.html}  \citep{Carrasco13} 
is used to identify catastrophic redshifts.
Photo-z results from these codes have been previously studied for DES in \citet{Sanchez14}. They use 
DES-SVA1 data trained with a spectroscopic redshift ($z_{sp}$) sample independent of the eBOSS data. 
\citet{Sanchez14} training sample contains 9000 $z_{sp}$ from various spectroscopic surveys. 
%the most secure $z_{sp}$ of OzDES , SDSS, 6dF, ATLAS, GAMA, SNLS, CBD, VVDS, VIPERS, XMM, ACES, DEEP2,  2dFGRS, UDSz. 
ANNz2 includes a wheighting of the galaxies during the training procedure which is a function
of magnitude, color and spectroscopic redshifts to correct for the sample difference between the
photometric and spectroscopic data. 

To separate stars and galaxies we use a combination of SExtractor parameters
combined to have an optimum purity/completness, hereafter modest\_class.
Modest\_class stars are defined in Table \ref{tab:modest_class}\\
%- FLAGS\_I $\geq$ 3 \& \\
%--- (CLASS\_STAR\_I $>$ 0.3 \& MAG\_AUTO\_I $<$ 18.0) \& MAG\_PSF\_I $<$ 30.0 \\
%--- OR \\
%--- (SPREAD\_MODEL\_I $+$ 3*SPREADERR\_MODEL\_I) $<$ 0.003 \& (SPREAD\_MODEL\_I $+$ 3*SPREADERR\_MODEL\_I) $>$ -0.003 \\
\begin{table*}
\caption{Modest\_class stars. Note that we add selection criterion at $g<23$ for the modest\_class classification. Magnitudes are the SExtractor \citep{Bertin96}.}
\[ 
\left \{
  \begin{tabular}{cccccccc}
  FLAGS\_I $\geq$ 3 & CLASS\_STAR\_I $>$ 0.3 & MAG\_AUTO\_I $<$ 18.0) \quad  MAG\_PSF\_I $<$ 30.0 \\ 
  OR &&& \\
  FLAGS\_I $\geq$ 3 & (SPREAD\_MODEL\_I $+$ 3*SPREADERR\_MODEL\_I) $<$ 0.003 & (SPREAD\_MODEL\_I $+$ 3*SPREADERR\_MODEL\_I) $>$ -0.003 & \\
  \end{tabular}
\right \}
\]
\label{tab:modest_class}
\end{table*}

Note that we add selection criterion at $g<23$ for the modest\_class classification. 

\subsection{eBOSS ELG spectroscopic targets}
\label{sec:eBOSS_ELG_SP_T}
%\textcolor{blue}{Can we add a summary of the spectroscopic data available in this field : VVDS, VIPERS before wew explain what eBOSS bring in terms of different redshift, resolution ...}

\begin{figure*}
\advance\leftskip-1.7cm
\includegraphics[width=1.2\textwidth,height=0.3\textwidth]{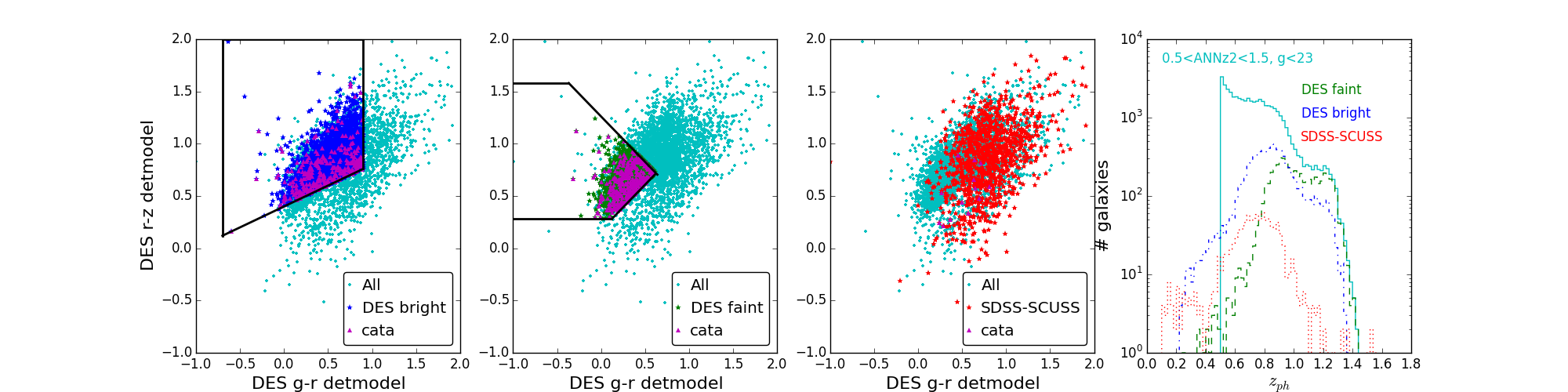}
%\resizebox{\columnwidth}{!}{\includegraphics{Photoz_distribution_eBOSStargets_SNfields.pdf}}
\caption{The three panels from the left-hand side show the DES detmodel grz colors of the eBOSS plates,
showing the ELG selection functions.
The magenta triangles labeled `cata' shows catastrophic redshifts as defined in section \ref{sec:photoz}.
The furthest right planel shows the
photometric redshift distribution of the eBOSS targets selected with the DES-SVA1. Photo-z are computed in using ANNz2 on the DES SVA1 data for the DES selections. For SDSS-SCUSS selection, we used CFHTLS photo-z which we matched with SDSS photometry. The median uncertainties on colors $g-r$ and $r-z$ for the different selections is less than 4\%.}
\label{fig:test_plates}
\end{figure*}

We used three tiles from the SVA1 data on CFHTLS-W1 which we observed in 8 eBOSS plates. 
With a one hour exposure we reached a total of 5705 spectra. 
We investigated three different target selection schemes, see Table \ref{tab:selections}. 
%\begin{enumerate} 
%\item  DES bright: $20.5<g<22.8 \quad\&\quad -0.7<g-r<0.9 \quad\&\quad 0<r-z<2 \quad\&\quad r-z>0.4*(g-r)+0.4$ (DES) 
%\item  DES faint: $g>20.45$ \quad\&\quad $r<22.8 \quad\&\quad 0.28<r-z<1.58 \quad\&\quad g-r<1.15*(r-z)-0.2 \quad\&\quad g-r<1.45-1.15*(r-z)$ (DES)
%\item  URI: $eg<0.6 \quad\&\quad er<1 \quad\&\quad ei<0.4 \quad\&\quad 20<g<23 \quad\&\quad r<22.5 \quad\&\quad i<21.6 \quad\&\quad 21< U<22.5 \quad\&\quad r-i>0.7 \quad\&\quad i-u>-3.5*(r-i)+0.7$ (SDSS/SCUSS)\\
%or GRI: $eg<0.6 \quad\&\quad er<1 \quad\&\quad ei<0.4 \quad\&\quad 21<=g<22.5 \quad\&\quad r<22.5 \quad\&\quad i<21.6 \quad\&\quad g-r<0.8 \quad\&\quad r-i>0.8$ (SDSS)
%%%%%% \item  DES\_faint: $20.5<g<23 \quad\&\quad -0.7<g-r<0.9 \quad\&\quad 0<r-z<2 \quad\&\quad r-z>0.6*(g-r)+0.22$ 
%\end{enumerate}
\begin{table*}
\caption{The three eBOSS ELG selections. $eg,er,ei,ez$ are photometric uncertainties of the g,r,i,z bands.Magnitudes are the SExtractor \citep{Bertin96} detmodel DES magnitudes for the bright and faint selections and SDSS/SCUSS model magnitudes for the SDSS-SCUSS selections.}
\[ 
\left \{
  \begin{tabular}{cccccccc}
  DES bright: & 20.5$<$g$<$22.8 & -0.7$<$g-r$<$0.9 & 0$<$r-z$<$2 & r-z$>$0.4*(g-r)+0.4 &&& \\
  DES faint: & $g>20.45$ & r$<$22.8 & $0.28<r-z<1.58$ & g-r$<$1.15*(r-z)-0.2 & g-r$<$1.45-1.15*(r-z) && \\
  \multirow{2}{*}{SDSS-SCUSS:} & \multirow{2}{*}{\shortstack[l]{eg$<$0.6 \\ er$<$1 \\ ei$<$0.4}}& 20$<$g$<$23 & r$<$22.5 & i$<$21.6 & 21$<$U$<$22.5  & r-i$>$0.7 & i-u$>$-3.5*(r-i)+0.7\\
  			  & & 21$<$g$<$22.5 & r$<$22.5 & i$<$21.6 & g-r$<$0.8 & r-i$>$0.8 &\\
  \end{tabular}
\right \}
\]
\label{tab:selections}
\end{table*}
SDSS-SCUSS targets are dispached over 51deg$^2$. DES bright and faint targets are dispached over 9.2deg$^2$.
SDSS-SCUSS is a combination of an SDSS only and a SDSS-SCUSS target selection as can be seen in Table \ref{tab:selections}.
The DES faint selection has been optimized to reach redshifts between 0.7 and 1.5. This latter selection
has been designed for the DESI survey and reaches higher redshifts galaxies than eBOSS is aiming at \citep{Schlegel11}.
eBOSS is aiming at galaxies between a redshift of 0.6 and 1.2. Higher redshifts will be explored using AGNs.
This papers studies the DES ELG target selections for eBOSS. The two other selections
using SDSS and SCUSS data will be presented in our companion paper \citet{Comparat15,Delubac15,Raichoor15}.
On the DES selections we apply the star-galaxy separation explained in section \label{sec:test_plates}.
using a combination of SExtractor output optimized for the DES data. We note that we do not
expect much contamination by stars when designing ELG target selection as shown in \citet{Adelberger04}. We remove the
fake detections by applying selection criteria in g,r,z DES bands of $MAG\_APERTURE - MAG\_DETMODEL < 2$. 
Figure \ref{fig:test_plates} shows the photometric redshifts distribution in cyan solid line of the galaxies in the DES 9.2 deg$^2$ field
used to optimise the eBOSS target selections. The cyan solid line shows all galaxies at $g<23$ with
a photometric redshift between 0.5 and 1.5. We show two of the DES based
target selections in dash-dotted blue and dashed green lines.
The red dotted line show an SDSS-based selection which we name SDSS-SCUSS, detailed in the next section.
The magenta triangles show the outliers. We note that there is color space where there seem to be a higher percentage of outliers,
especially for the DESI selection.
In section \ref{subsec:outliers_tree}, we present a first attempt to find color-magnitude boxes where 
we find a higher percentage of outliers. 
%%%%%%%%%%%%%%%%%%%%%%%%%%%%%%%%%%%%%%%%%%%%%%%%%%%%%%%%%%%%%%%%%%%%%%%%%%%%%%%%%%%%%%%%%%%%%%%%%%
%%% routine colcol in spec_phot_ELGtest_ColorCuts_nov2014_DES.py
%%%%%%%%%%%%%%%%%%%%%%%%%%%%%%%%%%%%%%%%%%%%%%%%%%%%%%%%%%%%%%%%%%%%%%%%%%%%%%%%%%%%%%%%%%%%%%%%%%

The eBOSS ELG observations are presented in Table \ref{tab:targets}.
We show the number of targets selected and observed in the eBOSS test plates. 
SDSS-SCUSS do not show the total number of targets observed
but the one for which we find a match with DES photometry. For representative statistics about the SDSS and SCUSS
selections, please refer to \citet{Comparat15,Delubac15,Raichoor15}.
We show the percentage of $z_{sp}$ with a secure redshift
``secure" for which we find at least two lines with a low signal-to-noise detection, or one line and 
a 10 $\sigma$ continuum detection for the redshift measurement. The ``1line"
were measured from a single line with at least 3$\sigma$ detection without continuum information. 
They have a higher failure rate since a
line confusion can happen between $[H\alpha]$ and $[OII]$. 
The ``unknown" are spectra for which we couldn't find
a redshift. `$0.6<z<1.2$' shows the percentage of targets with secure
redshifts in the desired redshift range: $0.6<z<1.2$. `$0.6<z<1.2*$' includes the `1line' $z_{sp}$ in the percentage
of targets with secure redshifts.
$\bar{z}$ and $<[OII]>$ are respectively the mean eBOSS $z_{sp}$ and $[OII]$ flux
using secure redshifts. Nstars is the number of stars. 

DES gives the highest success rate with 72\% of very secure redshift, 12\% of one line detected redshift, and 15\%
of non-identified redshift. DES faint selection has a slightly lower success rate of 68\% of very secure redshifts including 20\% of one line $z_{sp}$.
Table \ref{tab:targets} shows the success rate as a function of DES g-band magnitude. 
The DES faint selection has been designed to target fainter and higher
redshift galaxies which explains the slightly lower success rate when compared to DES bright selection.
In section \ref{sec:final_targets}, we apply the DES bright selection to the year one DES data and 
show the results of our systematics studies.
\begin{table*}
\caption{Number of targets for the four eBOSS selections. We use the DES `g' band to do magnitude selections.}
\begin{tabular}{ccccccc}
 &  & DES bright & DES faint & SDSS-SCUSS & DES bright$\cap$faint \\
\hline\hline
\multirow{4}{*}{}
	        &Selected & 953 & 445 & - & 220  \\
	        &Dens. Selected /deg$^2$& 69& 32 & - & 24  \\
		&Observed & 557 & 254 & 206 & 199 \\
		&secure(\%)& 88.0 & 85.0 & 77.2 & 87.4 \\
		&1line(\%) & 1.3 & 3.5 & 1.0 & 2.5 \\
$20.5<g<22$	&unknown(\%) & 10.8 & 11.4 & 21.8 & 10.1 \\
		&$0.6<z<1.2$(\%) & 60.9 & 66.5 & 58.3 & 71.4 \\
		&$0.6<z<1.2*$(\%) & 61.4 & 67.7 & 58.3 & 72.4 \\
		&$\bar{z}$ & 0.68 & 0.8 & 0.65 & 0.8 \\
		&$<[OII]>$ & 1.8 & 2.54 & 3.91 & 2.65\\ 
		& Nstars & 21 & 13 & 25 & 5 \\
\hline\hline
\multirow{4}{*}{}
		&Selected & 6762 & 7838 & - & 2158 \\ 
		&Dens. Selected /deg$^2$& 491& 570 & -  & 239  \\
		&Observed & 3103 & 2158 & 1049 & 1274 \\
		&secure(\%)& 70.6 & 64.0 & 74.5 & 67.1 \\
		&1line(\%) & 13.4 & 22.3 & 4.6 & 21.3 \\
$22<g<23$	&unknown(\%) & 16.0 & 13.6 & 21.0 & 11.6 \\
		&$0.6<z<1.2$(\%) & 64.8 & 55.7 & 62.7 & 60.4 \\
		&$0.6<z<1.2*$(\%) & 73.5 & 68.1 & 66.0 & 72.0\\
		&$\bar{z}$ & 0.83 & 0.88 & 0.71 & 0.9\\
		&$<[OII]>$ & 1.24 & 1.47 & 0.97 & 1.71\\
		& Nstars & 51 & 35 & 24 & 16\\
\hline\hline
\multirow{4}{*}{}
		&Selected & 7716 & 8283 & - & 2378\\
		&Dens. Selected /deg$^2$& 561& 602 & - & 264  \\
		&Observed & 3660 & 2412 & 1255 & 1473 \\
		&secure(\%)& 73.3 & 66.3 & 74.9 & 69.9 \\
		&1line(\%) & 11.6 & 20.4 & 4 & 18.7 \\
$20.5<g<23$	&unknown(\%) & 15.2 & 13.4 & 21.1 & 11.4 \\
		&$0.6<z<1.2$(\%) & 64.2 & 56.8 & 62 & 61.8 \\
		&$0.6<z<1.2*$(\%) & 71.6 & 68.0 & 64.7 & 72.0 \\
		&$\bar{z}$ & 0.8 & 0.87 & 0.7 & 0.88 \\
		&$<[OII]>$ & 1.34 & 1.61 & 1.47 & 1.87\\
		& Nstars & 72 & 48 & 49 & 21  
\end{tabular}
\label{tab:targets}
\end{table*}

\begin{figure}
\resizebox{\columnwidth}{!}{\includegraphics{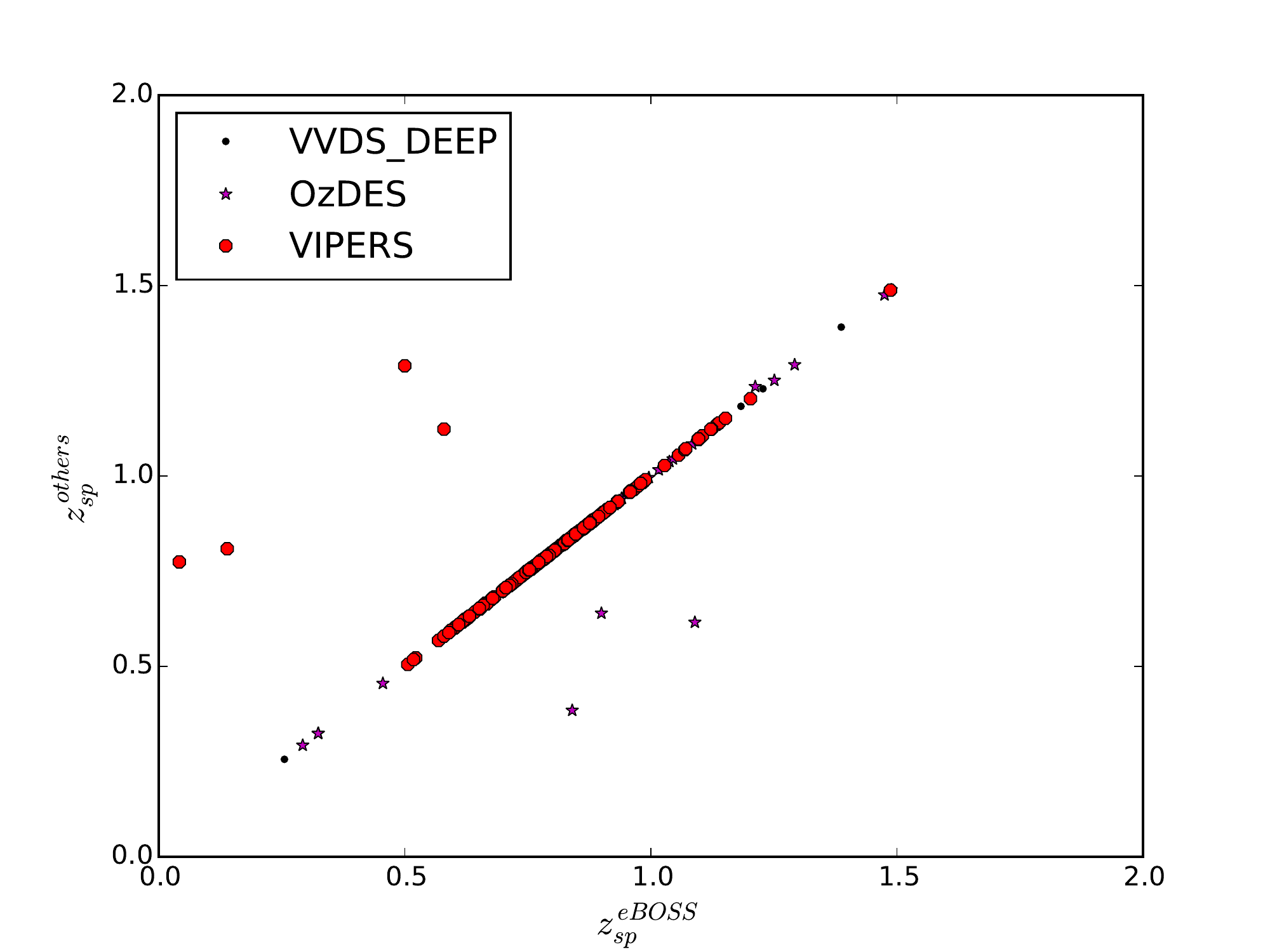}}
\caption{Spectroscopic redshift from eBOSS compared to other $z_{sp}$ surveys. We use only the secure $z_{sp}$.}
\label{fig:other_specz}
\end{figure}

Figure \ref{fig:other_specz} shows a comparison between our spectroscopic redshift measurement with other 
spectroscopic redshift surveys 
available on the CFHT-LS field: VVDS\_DEEP \citep{LeFevre05}, DEEP2 \citep{Newman13}, 
GAMA \footnote{http://www.gama-survey.org}, VIPERS \citep{Garilli14}, OzDES \citep{Yuan15}.
We used only the most secure redshift for both eBOSS and other $z_{sp}$ surveys.  
We matched catalogues at less than one arcsec.

With the VIPERS survey, using the quality flags 3.1, 3.2, 3.5, 4.1, 4.2 and 4.5, we find 4 galaxies out of 243 with a discrepant
redshift. For OzDES, we found 4 galaxies with discrepant redshift out of 95. However, restricting to the flag 4 only, 
we found no discrepant redshift between OzDES and eBOSS. Similarly, we found no discrepant redshift 
with the 23 galaxies in VVDS. If we restrict the match between catalogues at less than 0.1 arcsec, we don't find any
disagreement between VIPERS and eBOSS for the flags 3 and 4 redshifts. 
\citet{Comparat15} in preparation gives a full detailed study of the eBOSS redshift measurement pipeline and tests and 
finds an agreement with VIPERS at less than 1\%. We conclude that eBOSS redshift measurement are very reliable.
% GAMA \footnote{http://www.gama-survey.org}
%Figure \ref{fig:other_specz} shows the discrepancies that we are facing
%in terms of automated spectroscopic redshift measurement pipeline.  
% SDSS \citep{Ahn14}
%UDS \footnote{http://www.nottingham.ac.uk/astronomy/UDS/UDSz/}
%SNLS \footnote{http://cfht.hawaii.edu/SNLS/}
%XMM \citep{Stalin10}
%UDS \footnote{http://www.nottingham.ac.uk/astronomy/UDS/UDSz/}
%%%%%%%%%%%%%%%%%%%%%%%%%%%%%%%%%%%%%%%%%%%%%%%%%%%%%%%%%%%%%%%%%%%%%%%%%%%%%%%%%%%%%%%%%%%%%%%%%%
%%%%%%%%%%%%%%%%%%%%%%%%%%%%%%%%%%%%%%%%%%%%%%%%%%%%%%%%%%%%%%%%%%%%%%%%%%%%%%%%%%%%%%%%%%%%%%%%%%

\subsection{eBOSS DES ELG selection characteristics}
\label{sec:final_targets}
%%%%%%%%%%%%%%%%%%%%%%%%%%%%%%%%%%%%%%%%%%%%%%%%%%%%%%%%%%%%%%%%%%%%%%%%%%%%%%%%%%%%%%%%%%%%%%%%%%
\begin{figure}
\resizebox{\hsize}{!}{\includegraphics{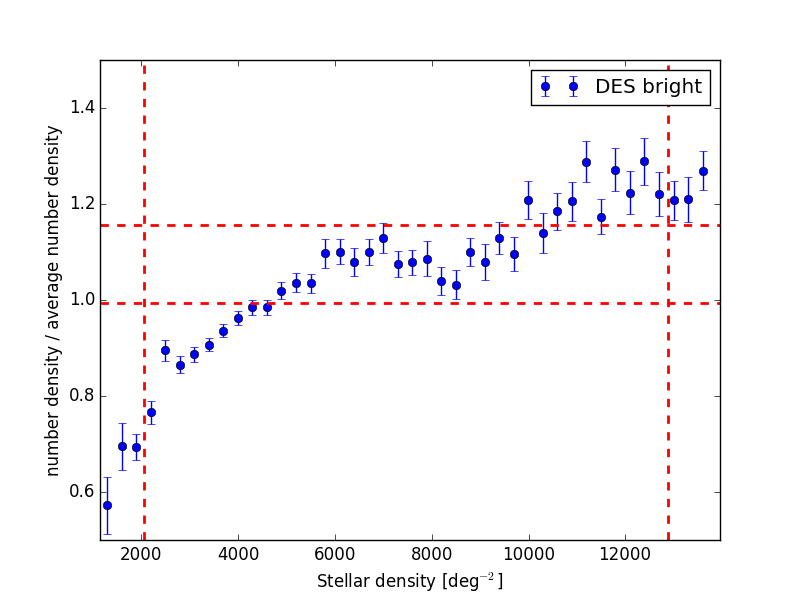}}
\caption{Density fluctuation of galaxies as a function of the stellar density.
The two horizontal and vertical red axes show respectively the 5 and 95\% of the star density distribution
and 15\% around the error weighted mean galaxy density fluctuation for the eBOSS target selection. 
The error weighted mean is computed using 94\% of the galaxy population, pruning from the extremes low and high
2\% of the density regions. Errorbars show the standard deviation of
the galaxy number density over the healpix pixels, that we used as weight in the mean calculation.} 
%\textcolor{blue}{Add prediction of the stellar density expected.}

\label{fig:sys_stellar}
\end{figure}
%%%%%%%%%%%%%%%%%%%%%%%%%%%%%%%%%%%%%%%%%%%%%%%%%%%%%%%%%%%%%%%%%%%%%%%%%%%%%%%%%%%%%%%%%%%%%%%%%%
%%%%%%%%%%%%%%%%%%%%%%%%%%%%%%%%%%%%%%%%%%%%%%%%%%%%%%%%%%%%%%%%%%%%%%%%%%%%%%%%%%%%%%%%%%%%%%%%%%
\begin{figure}
\resizebox{\hsize}{!}{\includegraphics{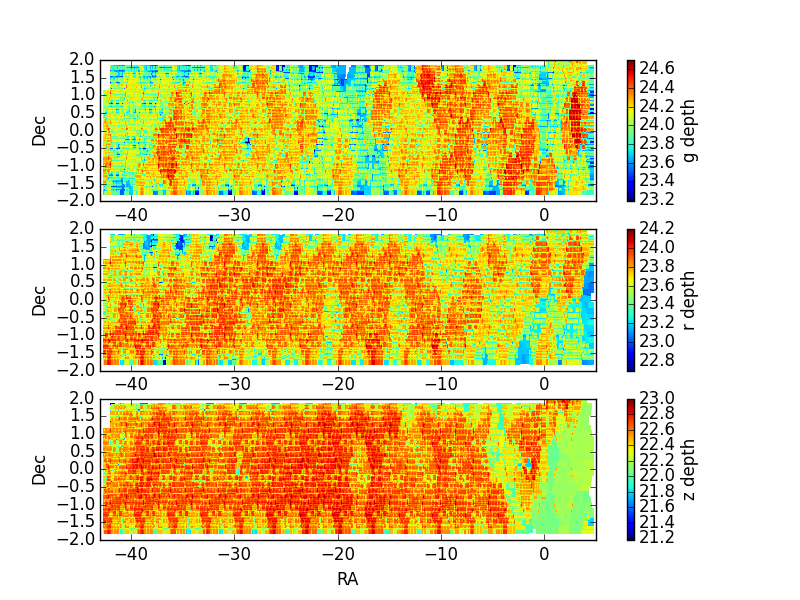}}
\caption{Depth of the DES year one data on Stripe 82 region in the g,r,z bands.}
\label{fig:y1a1_depth}
\end{figure}
%%%%%%%%%%%%%%%%%%%%%%%%%%%%%%%%%%%%%%%%%%%%%%%%%%%%%%%%%%%%%%%%%%%%%%%%%%%%%%%%%%%%%%%%%%%%%%%%%%
As mentionned in the introduction, DES and eBOSS footprint overlap on about 500deg$^2$ in the Stripe82 region.
The early release of the DES year one data, hereafter Y1A1, is on the Stripe82 field with 152deg$^2$.
We use the Y1A1 data to look at possible systematics from the photometric selections.
In order to have reliable measurement of the galaxy power spectrum in eBOSS we need a density
variation of maximum 15\% over the whole survey area as discussed in \citet{Ross12} and \citet{Dawson15} in preparation. 
We use Healpix\footnote{http://healpy.readthedocs.org/en/latest/} to produce maps of
the eBOSS galaxy target selection and systematics which have the biggest impact on the power spectrum
measurement such as stellar density, Galactic extinction, survey depth, airmass \citep{Ross12}.  
We used the DES bright selection for the galaxy density maps using a pixelisation of 6.87arcmin$^2$ (NSIDE=512).
Note that we add a selection criterion: $g<23$ for the modest\_class classification. We are aiming to test the impact of the
stars density variation with the eBOSS bright galaxies density variation. The density of stars at lower magnitude
will not impact our test. 
The number density of stars vary between 76 and 15000 with a mean of 5890 stars/deg$^{2}$. The mean number 
density of galaxies is 737 gal/deg$^{2}$ with variations between 19 to 2690 gal/deg$^{2}$. 
In Figure \ref{fig:sys_stellar}, we show the galaxy density fluctuation as a function
of the star density using modest\_class star classification. The galaxy density variation caused by
stars and their haloes is less than 15\% accross the the Y1A1 Stripe 82 survey area. We conclude that 
the photometric contamination of star haloes does not have a significant impact on our target selection.
Similarly, in Figures \ref{fig:sys_eboss}, we looked at the variation of target density as a function of depth, Galactic extinction, 
airmass accross the Y1A1 Stripe82 and conclude that our target selection will be within requirement accross the Stripe82 DES
footprint.

%For small areas where an unlucky combination of systematics would cause 
%density variation larger than 15\% around the mean, \citet{Ross12} suggest that this could be weighted out
%without much influence on the power spectrum measurement. 

\section{Photo-z from the eBOSS observations}
\label{sec:photoz}
%%%%%%%%%%%%%%%%%%%%%%%%%%%%%%%%%%%%%%%%%%%%%%%%%%%%%%%%%%%%%%%%%%%%%%%%%%%%%%%%%%%%%%%%%%%%%%%%%%
\begin{figure*}
\resizebox{\hsize}{!}{\includegraphics{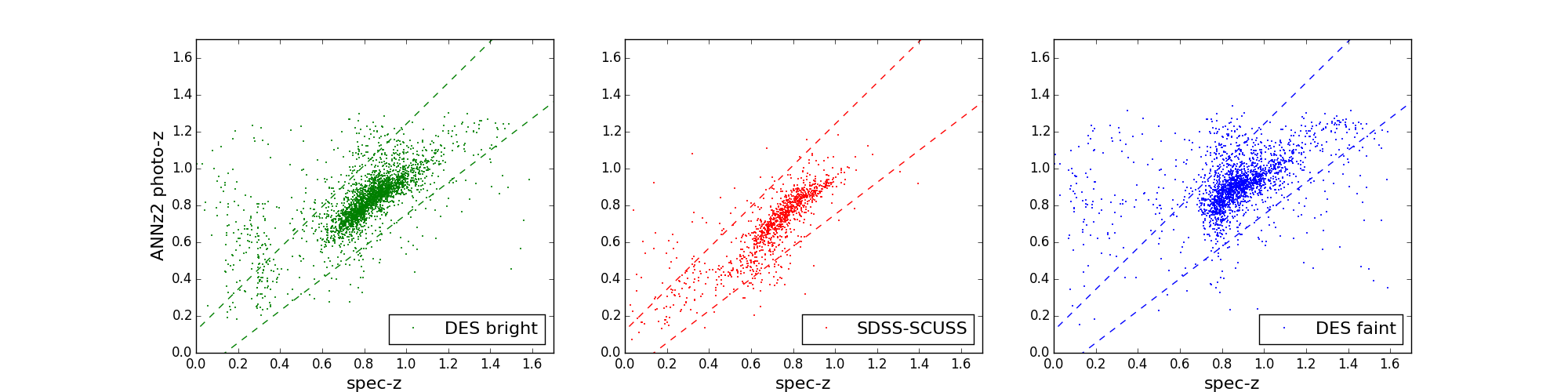}}
\resizebox{\hsize}{!}{\includegraphics{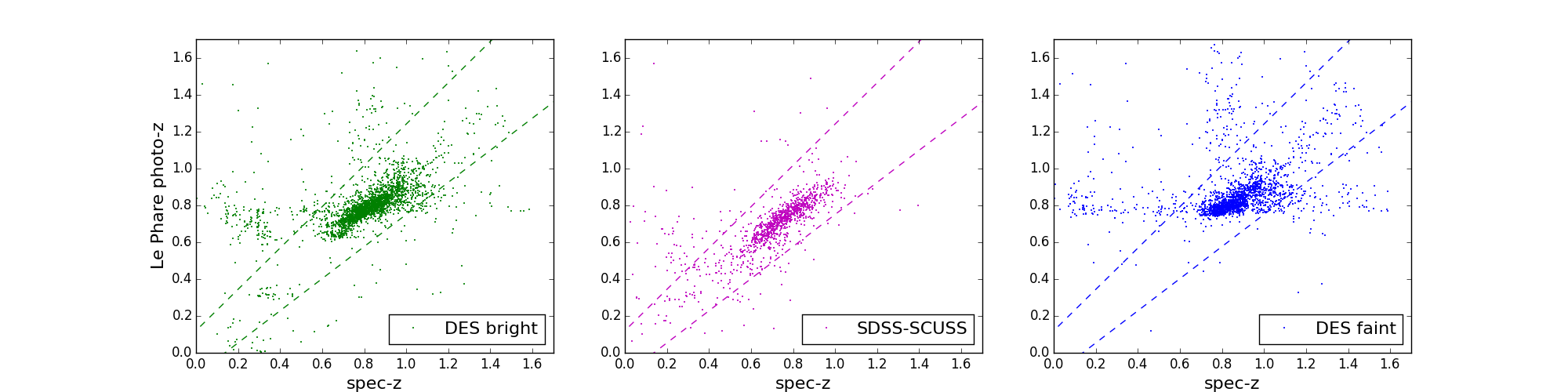}}
\caption{$z_{ph}$ vs $z_{sp}$ of the four eBOSS target selections. We use ANNz2 code for the top row and Le Phare for the bottom row,
on the five band DES-SVA1 photometry. The dashed line for the top and bottom panel shows the expected DES accuracy
of $|z_{ph}-z_{sp}|=\sigma_{DES}*(1+z_{sp})$ where $\sigma_{DES}=0.12$.}% The bottom row shows histograms of $z_{ph}$ and $z_{sp}$.}
\label{fig:eboss_photoz}
\end{figure*}
%%%%%%%%%%%%%%%%%%%%%%%%%%%%%%%%%%%%%%%%%%%%%%%%%%%%%%%%%%%%%%%%%%%%%%%%%%%%%%%%%%%%%%%%%%%%%%%%%%
\subsection{Photo-z of the four eBOSS selections}
We matched positions of the ANNz2 photometric redshifts catalogue \citep{Sanchez14} with positions of eBOSS 
targets with spectroscopic redshifts. 
We show photometric redshifts of the four eBOSS selections in Figure \ref{fig:eboss_photoz}, using
a comparison with the secure redshifts only. Note that \citet{Sanchez14} dealt with the full DES galaxy 
population, while the eBOSS emission-line galaxy sample considered here is a subset of the galaxy population 
for which photo-z's are prone to larger errors.
The SDSS-SCUSS selection has better photometric redshifts than the DES bright and faint selections as shown in Table \ref{tab:photz_targets}.
SDSS-SCUSS are targetted using the SDSS photometry which is shallower than the DES photometry.
SDSS-SCUSS targets are redder in g-r than DES targets as shown in Figure \ref{fig:test_plates}.
Galaxies with a large difference between g and r have a stronger Balmer or 4000A break making the
redshift measurement easier for $z<1$ galaxies. Indeed the SDSS-SCUSS selection has
criteria in r-i and u-i/g-r excluding the selection of galaxies with SED-power-law, which are difficult to
locate in redshift space. 
DES selections will then have a higher catastrophic rate as shown in Figures \ref{fig:test_plates}, 
\ref{fig:eboss_photoz} and Table \ref{tab:photz_targets}. Table \ref{tab:photz_targets} 
shows the mean and median $z_{ph}$, standard deviation, NMAD and outliers fraction 
of the $z_{ph}-z_{sp}$ distribution as a function of g-band magnitude for the four selections. 
NMAD is the normalised median absolute deviation defined as $1.48.median|(z_{ph}-z_{sp})/(1+z_{sp})|$.
NMAD is a calculation of the $z_{ph}-z_{sp}$ dispersion reducing the deviations induced by
outliers as explained in \citet{Ilbert09}.

For comparison we also obtained photometric redshifts with the Le Phare code \citep{Arnouts99,Ilbert06}.
We used the template library developed for the COSMOS observations in
\citet{Ilbert09}. It has 31 templates from elliptical to starbursts galaxies.
We apply extinction to the bluest templates using the Calzetti law \citep{Calzetti00}
with extinction values of $E(B-V)=[0.1,0.2,0.3]$. We enable the Le Phare in-built redshift prior
calibrated with the VVDS observations \citep{LeFevre05}. DES observations have the same depth as  
VVDS which justify the use of this prior. 
Figure \ref{fig:eboss_photoz} shows Le Phare and ANNz2 photometric redshifts as a function of the eBOSS spectroscopic redshifts. 
%We define outliers as galaxies for which the $z_{ph}$ is outside 2$\sigma_{exp}$ from the $z_{sp}$, where
%$\sigma_{exp}=0.12$, the expected DES $z_{ph}$ accuracy. The standard deviation is calculated using galaxies
%inside with $z_{ph}$ inside the 2$\sigma_{exp}$ from their $z_{sp}$. 
ANNz2 and Le Phare photometric redshifts have similar performances, although Le Phare has a tendancy to aggregate galaxies at 
$z_{ph}~0.8$. This is a feature caused by the discretisation of the redshift-template space encoded in most template fitting photometric
redshifts software. 
%%%%%%%%%%%%%%%%%%%%%%%%%%%%%%%%%%%%%%%%%%%%%%%%%%%%%%%%%%%%%%%%%%%%%%%%%%%%%%%%%%%%%%%%%%%%%%%%%%
\begin{table*}
\caption{ANNz2 photometric redshifts results for the four eBOSS selections. The DESz selection corresponds to the DES targets
selected at $0.6<z_{sp}<1.2$.}
\begin{tabular}{cccccccccc}
 &  & DES bright & DES faint & SDSS-SCUSS & DESz bright & DESz faint & SDSS-SCUSSz \\
\hline\hline
\multirow{4}{*}{}
		& $\bar{z_{ph}} $ & 0.72 & 0.8 & 0.63 & 0.8 & 0.83 & 0.71\\ 
$20.5<g<22$	&median($z_{ph}$) & 0.74 & 0.82 & 0.67 & 0.79 & 0.85 & 0.72\\ 
		&$\sigma[z_{ph}-z_{sp}]$ & 0.21 & 0.3 & 0.12 & 0.14 & 0.2 & 0.09\\ 
		&NMAD[$z_{ph}-z_{sp}$] & 0.07 & 0.11 & 0.04 & 0.05 & 0.08 & 0.03 \\
		&outliers & 101 & 71 & 10 & 43 & 40 & 5 \\

\hline\hline
\multirow{4}{*}{}
		& $\bar{z_{ph}}$ & 0.85 & 0.93 & 0.7 & 0.85 & 0.92 & 0.75 \\
$22<g<23$	&median($z_{ph}$) & 0.85 & 0.92 & 0.72 & 0.85 & 0.92 & 0.75 \\
		&$\sigma[z_{ph}-z_{sp}]$ & 0.15 & 0.22 & 0.11 & 0.1 & 0.13 & 0.09 \\
		&NMAD[$z_{ph}-z_{sp}$] & 0.04 & 0.05 & 0.04 & 0.03 & 0.04 & 0.03 \\
		&outliers & 187 & 223 & 37 & 83 & 104 & 21\\ 
\hline\hline
\multirow{4}{*}{}
		&$\bar{z_{ph}}$ & 0.83 & 0.91 & 0.69 & 0.84 & 0.91 & 0.74 \\ 
$20.5<g<23$	&median($z_{ph}$) & 0.84 & 0.92 & 0.72 & 0.84 & 0.91 & 0.75 \\
		&$\sigma[z_{ph}-z_{sp}]$ & 0.17 & 0.23 & 0.11 & 0.11 & 0.14 & 0.09\\ 
		&NMAD[$z_{ph}-z_{sp}$] & 0.04 & 0.06 & 0.04 & 0.03 & 0.05 & 0.03 \\
		&outliers & 288 & 294 & 47 & 126 & 144 & 26 
\end{tabular}
\label{tab:photz_targets}
\end{table*}

\subsection{Removing catastrophics: template fitting vs neural network}
\label{subsec:TF_NN}
\begin{figure}
\resizebox{\hsize}{!}{\includegraphics{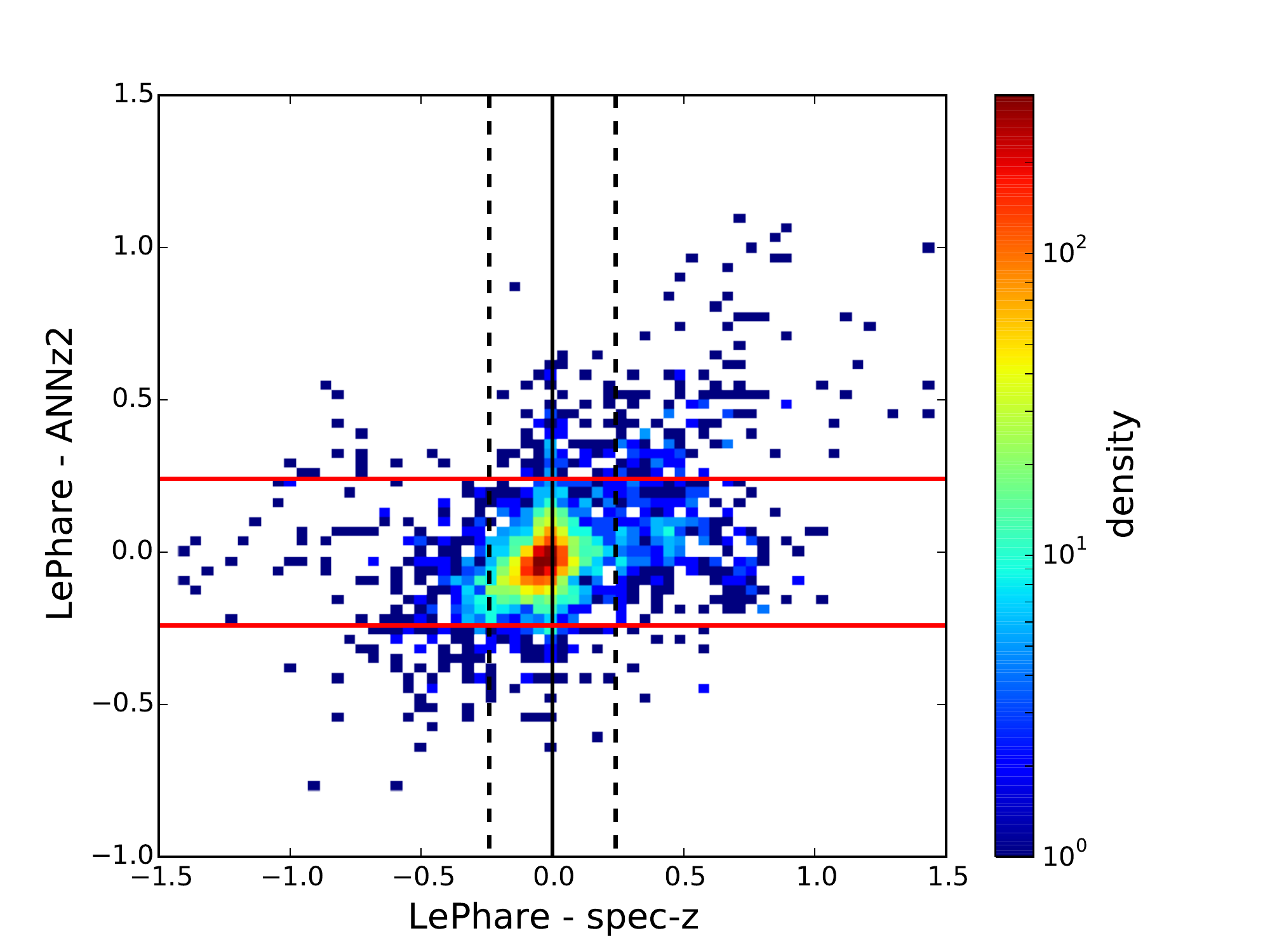}}
\resizebox{\hsize}{!}{\includegraphics{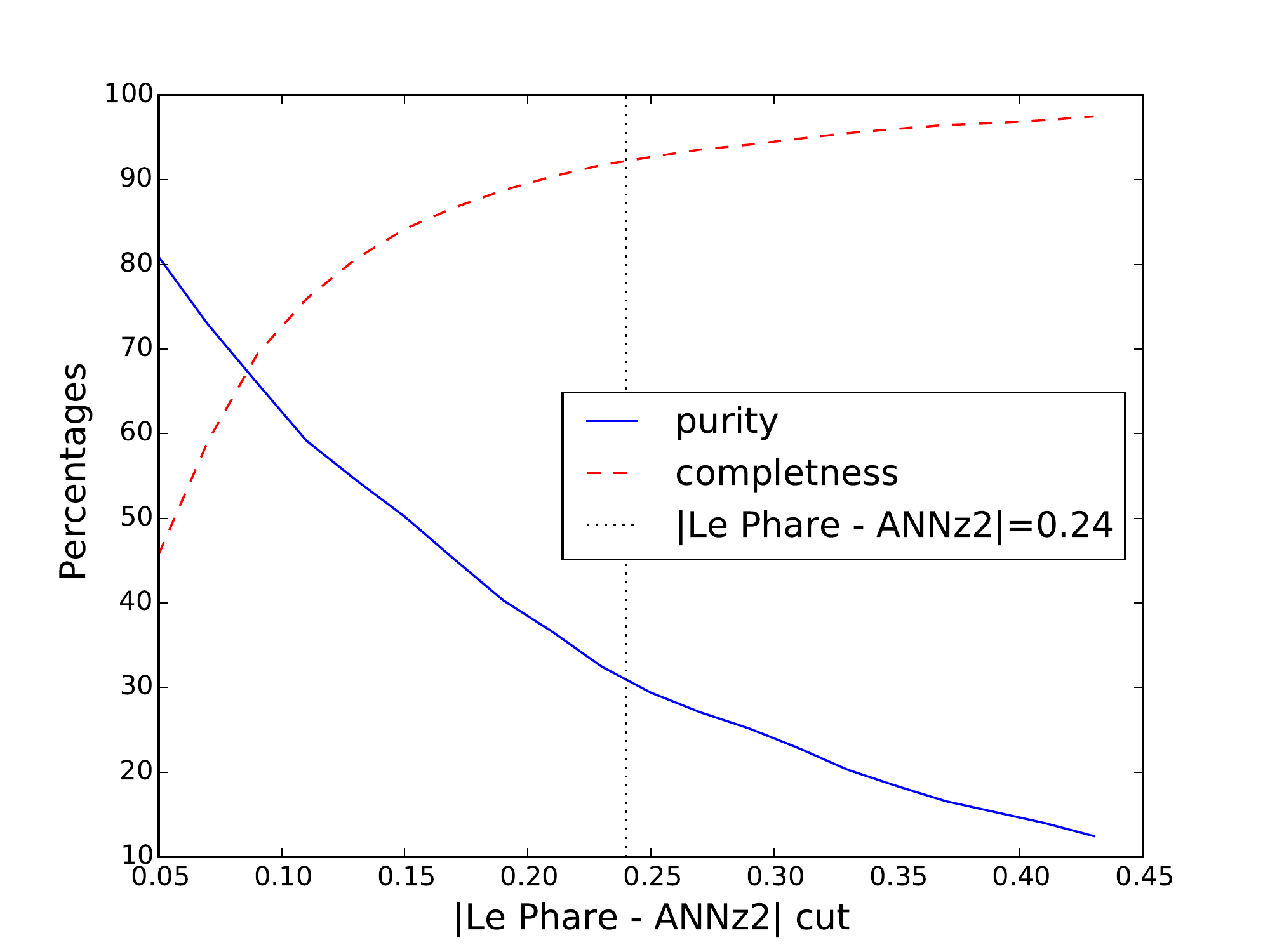}}
\caption{Top panel shows the $z_{ph}$ improvement while doing a selection criterion on the difference between template fitting and machine learning $z_{ph}$.
The bottom panel shows the purity-completness while changing the selection criterion. The bottom panel shows how well this method would perform.}
\label{fig:cata_lp_ann}
\end{figure}

In the context of the DES survey, we investigate possible ways to remove catastrophic
redshifts. ELG are the most difficult galaxy population to find an accurate $z_{ph}$. 
ELG are the source of the high percentage of outliers. We take advantage of the eBOSS sample to
look at possible ways of calibrating and removing a part of the outliers fraction.
A first solution has been proposed in \citet{Newman13b}. Comparing
template fitting methods and machine learning helps at pruning outliers.
Figure \ref{fig:cata_lp_ann} top panel shows an example of this method with the DES-eBOSS data.  
Top panel of Figure \ref{fig:cata_lp_ann} shows the density
of $Le Phare - z_{sp}$ versus $LePhare - ANNz2$. Accurate photometric redshifts are located inside the black lines.
The red lines show a possible template fitting vs machine learning criterion at $|ANNz2 - LePhare| = 2*\sigma_{DES}$ where$\sigma_{DES}=0.12$
is the expected accuracy for the DES survey. Selecting the galaxies inside the
red lines is a possible way to prune from outliers. It will however remove some galaxies for which
we have a good photometric redshift. There is a tradeoff between removing more outliers in moving the red lines 
and not loosing too many galaxies for the clustering analysis.     
The bottom panel of Figure \ref{fig:cata_lp_ann} shows the purity and completness trend while
changing the 2*$\sigma_{DES}$ value to the value of the x-axis. 
The purity shows the percentage of catastrophic redshifts that the selection criteria 
will remove. Catastrophic redshifts are defined as $|LePhare-z_{sp}|>2\sigma_{DES}$. The completness
shows the percentage of galaxies left in the sample. A selection criteron at 2$\sigma_{DES}$ removes about 30\% of outliers
and leaves 92\% of the galaxy sample. Following this idea, a more detailed work has also been pursued in \citet{Carrasco14}. 

%%%%%%%%%%%%%%%%%%%%%%%%%%%%%%%%%%%%%%%%%%%%%%%%%%%%%%%%%%%%%%%%%%%%%%%%%%%%%%%%%%%%%%%%%%%%%%%%%%
%%%%%%%%%%%%%%%%%%%%%%%%%%%%%%%%%%%%%%%%%%%%%%%%%%%%%%%%%%%%%%%%%%%%%%%%%%%%%%%%%%%%%%%%%%%%%%%%%%

\subsection{Removing catastrophics using Random Forest}
\label{subsec:outliers_tree}
\begin{figure}
\vbox{
\resizebox{\hsize}{!}{\includegraphics{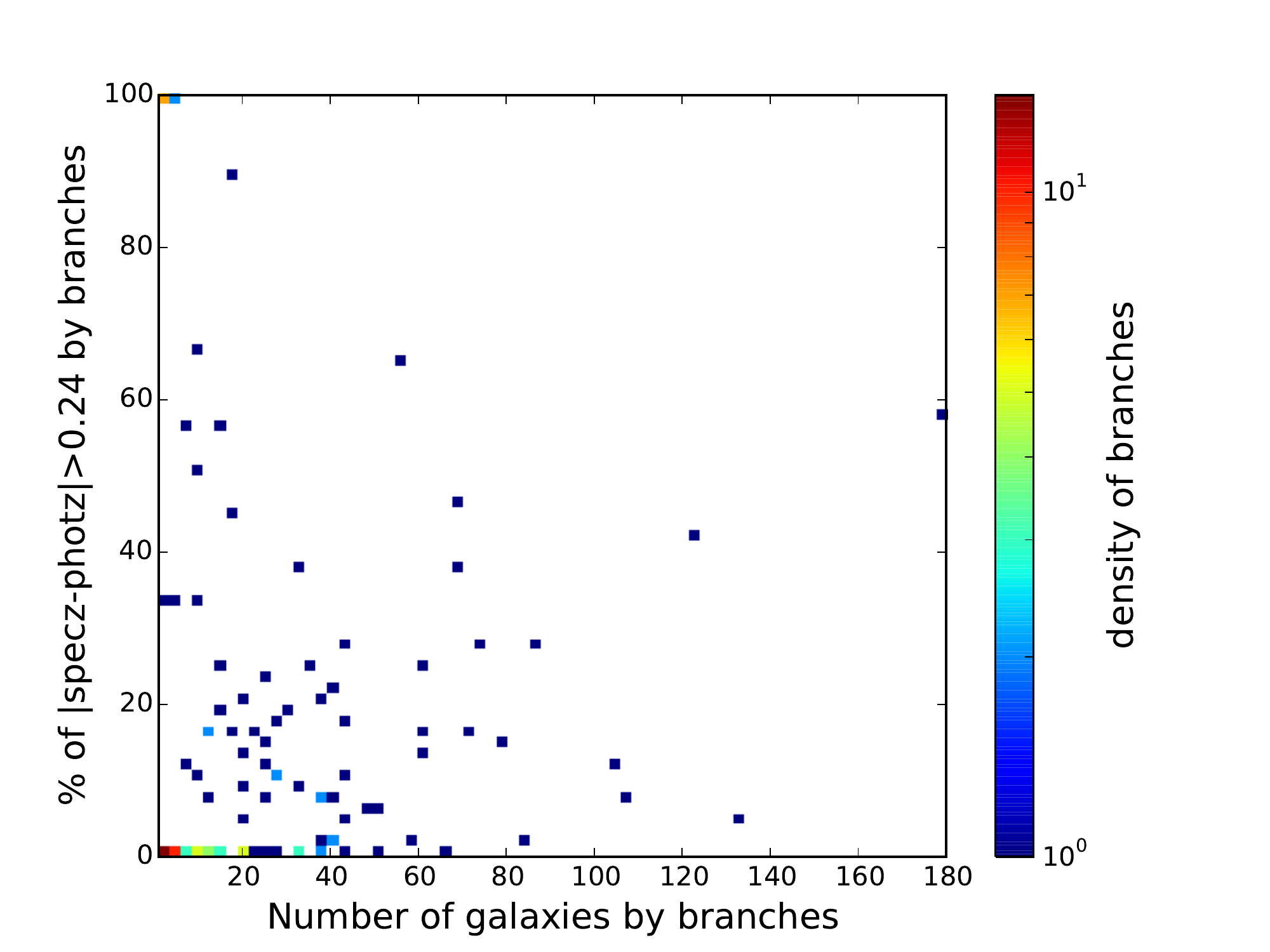}}
\resizebox{\hsize}{!}{\includegraphics{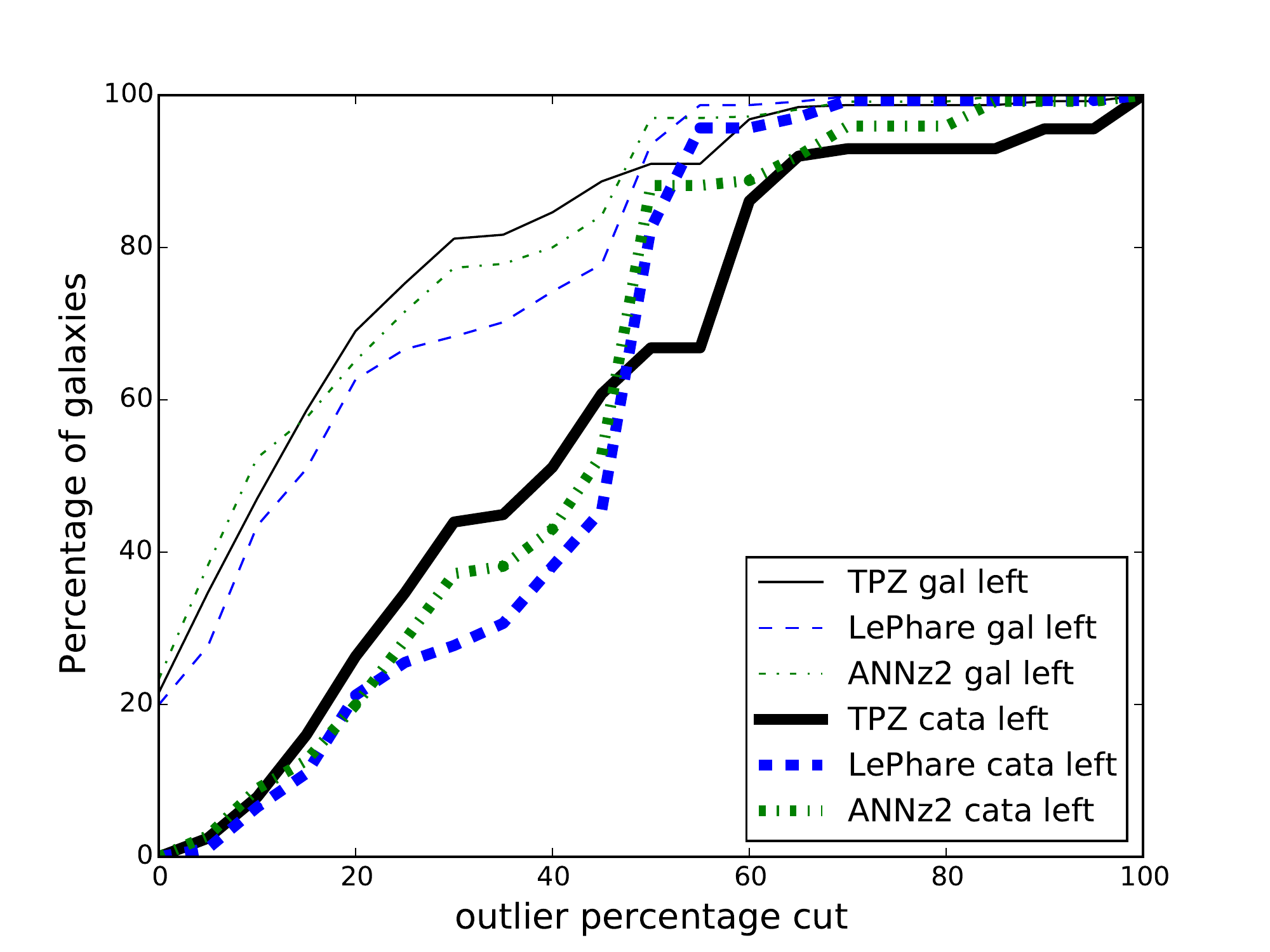}}
}
\caption{Top panel shows the number of galaxies as a function of the percentage of outliers by branches. Colors show the density of branches. 
Bottom panel shows the number of galaxies and catastrophic redshift left as a function of the outlier selection criterion applied on the branches.}
\label{fig:cata_tree}
\end{figure}

We investigated a new method to remove outliers using the TPZ code  which is part of the MLZ framework \citep{Carrasco13}. 
TPZ is a random forest algorithm which, in this case, divides the color space in branches, forming multi-dimensional color boxes. 
We used a random half of the eBOSS spec-z as a training set and the other half as testing set.
We give TPZ the DES magnitudes and colors.
We looked at the percentage of catastrophics for each of the branches. The top panel of 
Figure \ref{fig:cata_tree} shows the number of galaxies as a function of the number of catastrophic redshifts
for each of the branches. The density corresponds at the number of branches. We observe that
some branches have a high percentage of catastrophic redshift for a high number of galaxies in the box. 
Same as in section \ref{subsec:TF_NN}, there is a tradeoff between the percentage of outliers one can remove
and the percentage of galaxies which are left in the sample.
The bottom panel of Figure \ref{fig:cata_tree} shows the purity and completness while using this method.
Lines show the number of galaxies left in the sample as a function of the percentage of outliers selection criterion value used to trim
the branches. For example, if we exclude the boxes which have more than 40\% of outliers, we are left with about 85\% of
galaxies and 50\% of outliers, depending on which photometric redshift catalogue is used to define the outliers. 
In black solid, blue dashed, and green dashed-dotted, respectively we show the results for TPZ, Le Phare, and ANNz2. 
The axis on the right handside shows the percentage of galaxies left. 
We note that this method relies on having a spectroscopic sample (i.e., the eBOSS test plates), and that 
this sample then defines the good color boxes which may then be used in the future (if planned) to select 
eBOSS samples with fewer catastrophic outliers. This method could also be applied to decrease the outlier
fraction in the DES photo-z main sample, assuming a representative training set. 

%%%%%%%%%%%%%%%%%%%%%%%%%%%%%%%%%%%%%%%%%%%%%%%%%%%%%%%%%%%%%%%%%%%%%%%%%%%%%%%%%%%%%%%%%%%%%%%%%%
%%%%%%%%%%%%%%%%%%%%%%%%%%%%%%%%%%%%%%%%%%%%%%%%%%%%%%%%%%%%%%%%%%%%%%%%%%%%%%%%%%%%%%%%%%%%%%%%%%

\section{Clustering properties}
\label{sec:clustering}
In this section we measure the clustering properties of the DES galaxy target selections. 
We analyze the faint and bright selections separately and estimate the galaxy bias of 
these samples. Further clustering analyses for different eBOSS target selections will be 
presented in a separate work in preparation. 
Throughout this analysis we assume a flat $\Lambda CDM+\nu$ (one massive neutrino) cosmological model based on Planck 2013 + WMAP polarisation + $ACT/SPT$ + BAO, with a total matter density relative to critical $\Omega_{m}=0.307$ (\citep{2014A&A...571A..16P}).

Tables~\ref{tab:meanz} show the number of galaxies, mean density and redshift for the bright and faint samples considered 
in the clustering analysis. The spectroscopic catalogue has been selected using the true spectroscopic redshift in a given redshift range, 
while the \textit{ANNz2} and \textit{LePhare} catalogues are selected from their respective photometric redshift estimations, 
for galaxies in the range $0.6 < z_{ph} < 1.2$, but using their true redshift in order to compute distances and estimate the monopole of 
the two-point spatial correlation function (see below). The purity of each sample was calculated as the 
ratio of the number of true galaxies with spectroscopic redshifts in the given range and the number of galaxies in the spectroscopic sample. 
The mean redshift is calculated with the photometric redshifts, and in parenthesis, with the spectroscopic redshift of the same sample. 
By comparing the mean $z_{ph}$ and $z_{sp}$, we get an estimate of the bias for each sample.
The effective area of the footprint is 9.2 deg$^2$. These numbers were obtained after the catalog was pruned by the angular mask, as detailed below.

\begin{table}
\begin{tabular}{ccccc}
   & & bright  &   \\
\hline
   & Number (purity) & mean density & mean redshift  \\
\hline
$z_{sp}$ & 2613 (100\%) & 284.02 gal/sqdeg & 0.855 (0.855)  \\
ANNz2 & 2902 (86.66\%) & 315.43 gal/sqdeg & 0.866 (0.902)   \\
LePhare & 3038 (84.10\%) & 330.22 gal/sqdeg & 0.811 (0.894)  \\
\\
   & & faint &   \\\hline
   & Number (purity) & mean density & mean redshift  \\
\hline
$z_{sp}$ & 2139 (100\%) & 232.50 gal/sqdeg & 0.901 (0.901)  \\
ANNz2 & 2582 (79.43\%) & 280.65 gal/sqdeg & 0.928 (0.994)   \\
LePhare & 2662 (77.23\%) & 289.35 gal/sqdeg & 0.841 (1.004)  \\
\end{tabular}
\caption{Properties for the bright and faint sample including the different photometric redshifts catalogues properties. 
The neural network redshift code seems to have a better performance.
ANNz2 code produces counts and number densities that more closely resemble those from the 
sample with spectroscopic redshifts.}
\label{tab:meanz}
\end{table}

The redshift distribution for the spectroscopic sample is shown in Figure~\ref{fig:eboss_photoz}.
%We estimate the monopole of two-point spatial correlation function using the Landy-Szalay estimator under the fiducial cosmology, correcting from the integral constraint.
\subsection{Random fields}
 We use the Mangle mask of the DES survey (Figure~\ref{fig:grz_depth_eboss}) in the observed field to create a random catalog, 
sampling the footprint with the same depth and angular distribution than the ELG selections. We use as reference observations the g band,  
where the  magnitude limit is imposed to create the target selection (see target selection in \ref{sec:eBOSS_ELG_SP_T}). 
The magnitude limit distribution for the area analysed is shown in the first panel of Figure~\ref{fig:grz_depth_eboss} in units of 
magnitude in a 2 arcseconds aperture at $10\sigma$.

We compute the random sample considering the different magnitude limits of the area, applying a limiting aperture magnitude 
selection in the mask at $23.4 < mag < 25.8$. The final area mounts to 9.2 deg$^2$.

As we observe different galaxy densities depending on the survey depths, we take this information into account 
when generating the random catalogs.
In order to generate non-uniform random catalogs according to DES depths we have
applied the following methodology:
\begin{enumerate}
\item Create a uniform random catalog according to the galaxy catalog angular footprint.
\item Assign the magnitude limit in each position of the galaxy catalog and in the random uniform catalog
\item Using the Mangle mask, retrieve the information about the area and magnitude limit of each of the polygons into the footprint.
\item Create a histogram of magnitudes limits, in our case, 10 bins from 23.4 to 25.8
\item Count the number of galaxies in each magnitude limit bin (magL)
\item Build the density distribution in each magL as the number of galaxies over the
area in the given bin and generate the density(m) function, i.e., the density as a function of magL.
\item Create the Probability Function according to:
$$
P[i] = \frac{\int_{m_{min}^{i}}^{m_{max}^{i}}\rho(m)}{\int_{m_{min}^{tot}}^{m_{max}^{tot}}\rho(m)}
$$
where $m_{max}^{i}$ and $m_{min}^{i}$ are the maximum and the minimum values of magL in the bin i
and $m_{min}^{tot}$ and $m_{max}^{tot}$ are the initial and final magL according to the binning used. 
In our case, $m_{max}^{tot} = 25.8$ and $m_{min}^{tot} = 23.4$.

\item According to the magnitude limit where the random point lies, assign a probability to that point
\item Sample the probability function to reject/accept random points
\end{enumerate}

\begin{figure}
\resizebox{\hsize}{!}{\includegraphics{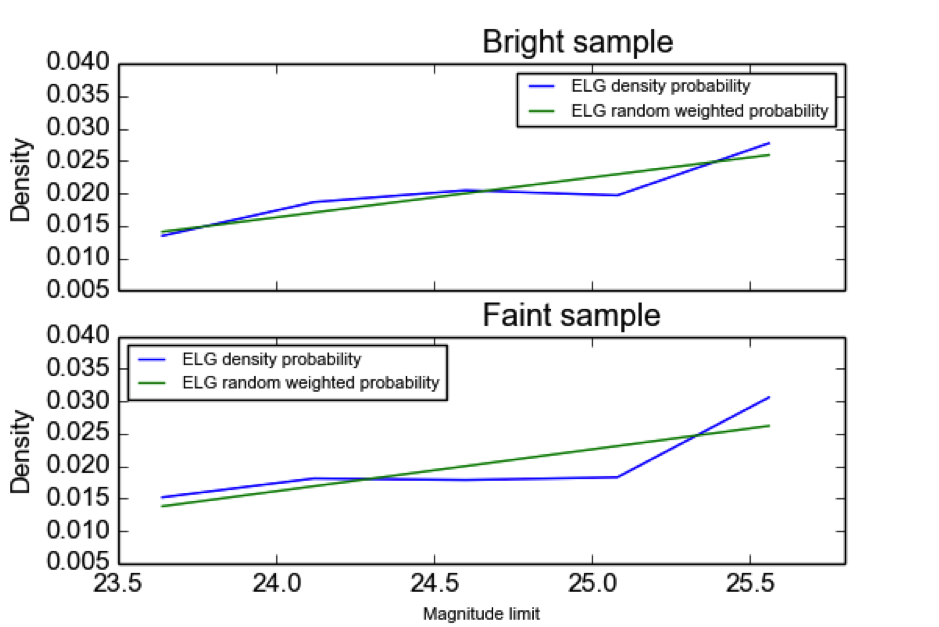}}
\caption{The blue lines show the density of targets as a function of magnitude limit in g band (given by the Mangle mask) 
for the bright sample (top panel) and faint sample (bottom panel). In green, the best fit to a first order polynomial, which, 
if we normalize to be 1 in its maximum, at $m=25.8$, represent the weight used in the random catalog to account for the different 
depths of the footprint. For some intermediate magnitude limits, there is an apparent decrease in density. 
This is definitely a variance effect, due to the small area observed in a very inhomogeneous footprint (see figure 2) and the small number of targets.} 
\label{fig:weight_random_clustering}
\end{figure}

In Figure~\ref{fig:weight_random_clustering}, we show the probability distribution as a function of magnitude limit 
according to the steps above, for both the bright and faint sample. 
This measurement is very limited by sample variance, nonetheless, we approximate the density distribution by a first 
order polynomial to assign  reject/acceptance probabilities as a function of magnitude limit for the random samples.
We find a mean error of 8\% for each of the ELG probability densities estimated using the variance cookbook given in ~\citep{Moster11} with similar survey configurations as COSMOS.  

%%%%%%%%%%%%%%%%%%%%%%%%%%%%%%%%%%%%%%%%%%%%%%%%%%%%%%%%%%%%%%%%%
%\caption{Density as a function of magnitude limit in the position (given by the mangle mask). On top, the best fit to a first order polynomial, which, normalize as it is equal to 1 at $m=25.8$, represent the weight for the random catalog.}

With this probability distribution as a function of the magnitude limit in the position of the footprint, we can now calculate 
the random catalog used throughout the following analysis. In both samples we calculate approximately $2.8e6$ random points.

%%%%
\subsection{Two-point spatial correlation function}
%%%%

We estimate the two-point spatial correlation function using the Landy \& Szalay \citep{1993ApJ...412...64L} estimator under 
the fiducial cosmology over scales $1 < s < 50$ Mpc/h using the CUTE code\footnote{http://members.ift.uam-csic.es/dmonge/CUTE.html} \citep{2012arXiv1210.1833A} 
and calculate the galaxy bias for the samples. In this section, we use the letter $s$ to refer to scales in redshift 
space for the calculation of the monopole, and $r_{p}$ for scales in real space, corrected from redshift space distortions, needed to calculate the bias. 

\begin{figure}
\resizebox{\hsize}{!}{\includegraphics{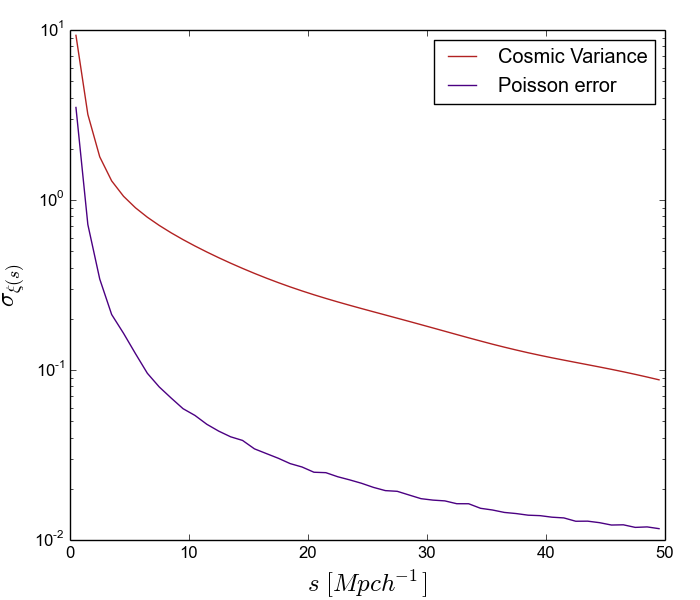}}
\caption{The cosmic variance error contribution in $\xi(s)$ compared to the Poisson unique contribution, used through the analysis. 
Clearly, the addition of this important term worsen the precision, so the errors given below in the analysis are underestimated compared to their true uncertainty.}
\label{fig:cosmic_var}
\end{figure}

During the analysis, we have only considered Poisson errors and therefore, the uncertainty in the results presented here is underestimated. 
To estimate how much we are underestimating it, we calculate the cosmic variance contribution to the error in the correlation function monopole 
using the analytic expression given in \citep{2012MNRAS.427.2146X}. The result can be seen in figure~\ref{fig:cosmic_var}. 
Clearly, the addition of the cosmic variance term in the error budget will worsen the precision of our results. 
A more thorough analysis of the clustering signal will demand the calculation of mocks catalogs for a more precise calculation of the covariance 
and in a bigger volume to have a significant clustering value.

%%%%%%%%%%%%%%%%%%%%%%%%%%%%%%%%%%%%%%%%%%%%%%%%%%%%%%%%%%%%%%
%%%%%%%%%%%%%%%%%%%%%%%%%%%%%%%%%%%%%%%%%%%%%%%%%%%%%%%%%%%%%%%%%%%%%%%%%%%%%%%%%%%%%%%%%%%%%%%%%%

We then fit $\xi(s)$ on small scales using a power-law
\begin{equation}
  \xi(s) = \left( \frac{s}{s_{0}} \right)^{-\gamma},
  \label{eqn:xis_powerlaw}
\end{equation}

In order to account for the finite volume of the sample, we measured the $RR$ terms up to maximum separations 
allowed by the volume and, following \citep{Roche:1998vz}, we estimated an integral constraint factor ($IC$) as
\begin{equation}
  IC = \frac{\sum_{i}^{} \xi(s_{i}) RR(s_{i})}{\sum_{i}^{} RR(s_{i})},
  \label{eqn:ic_xis}
\end{equation}
where $\xi(s)$ represents the model.

We fit the power law to the data subtracting the $IC$ from the model and allowing it to vary with the model parameters. 
We checked that this procedure is consistent with an iterative approach, in which we first fit a model to the original data, 
then use this model to estimate a correction via Eq.~\ref{eqn:ic_xis}, and apply this correction to the data, 
repeating the process to the new data until convergence 
is achieved. In our case, convergence was always reached in less than 20 iterations.
By fitting the model and $IC$ correction simultaneously, we avoid the need for correcting the data.

For the fit we used comoving scales in the range $1<s<20 h^{-1}$Mpc.
We considered different maximum scales between $10$ and $50 h^{-1}$Mpc to perform the fit and 
found that the results are not much affected by this scale.
Nonetheless $20 h^{-1}$Mpc was chosen because i) at the redshifts of interest the linear 
regime extends up to this scale, ii) for scales below $20 h^{-1}$ Mpc the amplitude of 
measured $\xi(s)$ is always one order of magnitude larger than our estimates of the $IC$. 
% and iii) the reduced $\chi^{2}$ for the fits is more consistent with unit for this maximum scale 
%compared to others.

%%%%%%%%%%%%%%%%%%%%%%%%%%%%%%%%%%%%%%%%%%%%%%%%%%%%%%%%%%%%%%%%%%%%%%%%%%%%%%%%%%%%%%%%%%%%%%%%%%
\begin{table*}
  \centering
  \begin{tabular}{cc|cccc}
    \hline
    Sample & Redshift selection & $s_{0}$ [$h^{-1}$ Mpc] & $\gamma$ & $IC$ & $\chi^{2}/{\rm dof}$ \\
    \hline
          & $z_{sp}$                            & $5.13^{+0.18}_{-0.19}$ & $1.313^{+0.059}_{-0.062}$ & 0.014 & 0.995 \\
          & ANNz2                            & $5.32^{+0.19}_{-0.21}$ & $1.255^{+0.060}_{-0.062}$ & 0.012 & 1.02 \\
    Faint & LePhare                          & $5.30^{+0.19}_{-0.20}$ & $1.265^{+0.059}_{-0.060}$ & 0.011 & 1.02 \\
          & ANNz2 $|$LePhare-ANNz2$|<0.24$   & $5.57^{+0.20}_{-0.21}$ & $1.279\pm0.062$ & 0.013 & 1.35 \\
          & LePhare $|$LePhare-ANNz2$|<0.24$ & $5.48^{+0.20}_{-0.22}$ & $1.260^{+0.063}_{-0.067}$ & 0.013 & 1.15 \\
    \hline
           & $z_{sp}$                            & $5.27^{+0.15}_{-0.16}$ & $1.264^{+0.045}_{-0.047}$ & 0.016 & 0.929 \\
           & ANNz2                            & $5.72^{+0.15}_{-0.16}$ & $1.261^{+0.044}_{-0.047}$ & 0.014 & 1.03 \\
    Bright & LePhare                          & $5.64^{+0.14}_{-0.15}$ & $1.260^{+0.043}_{-0.046}$ & 0.014 & 1.12 \\
           & ANNz2 $|$LePhare-ANNz2$|<0.24$   & $5.77^{+0.15}_{-0.16}$ & $1.257^{+0.045}_{-0.047}$ & 0.016 & 1.06 \\
           & LePhare $|$LePhare-ANNz2$|<0.24$ & $5.80^{+0.15}_{-0.17}$ & $1.230^{+0.044}_{-0.046}$ & 0.016 & 1.14 \\
    \hline
  \end{tabular}
  \caption{Power law fits for the monopole of the two-point correlation function of faint and bright DES target selection ELG samples. For each sample we show the impact on the parameters of the power law of selecting objects according to the different $z_{ph}$ methods considered. $1<s<20$ Mpc$/h$.}
  \label{tab:spatial_clustering}
\end{table*}
%%%%%%%%%%%%%%%%%%%%%%%%%%%%%%%%%%%%%%%%%%%%%%%%%%%%%%%%%%%%%%%%%%%%%%%%%%%%%%%%%%%%%%%%%%%%%%%%%%

Our results are shown in Table~\ref{tab:spatial_clustering}.
The $z_{sp}$ samples are selected with spectroscopic redshifts between $0.6$ and $1.2$. 
ANNz2 and LePhare are samples selected in the same range but using $z_{ph}$ estimates for each code; 
however we use their true redshifts to compute distances and $\xi(s)$.

\begin{figure}
\resizebox{\hsize}{!}{\includegraphics{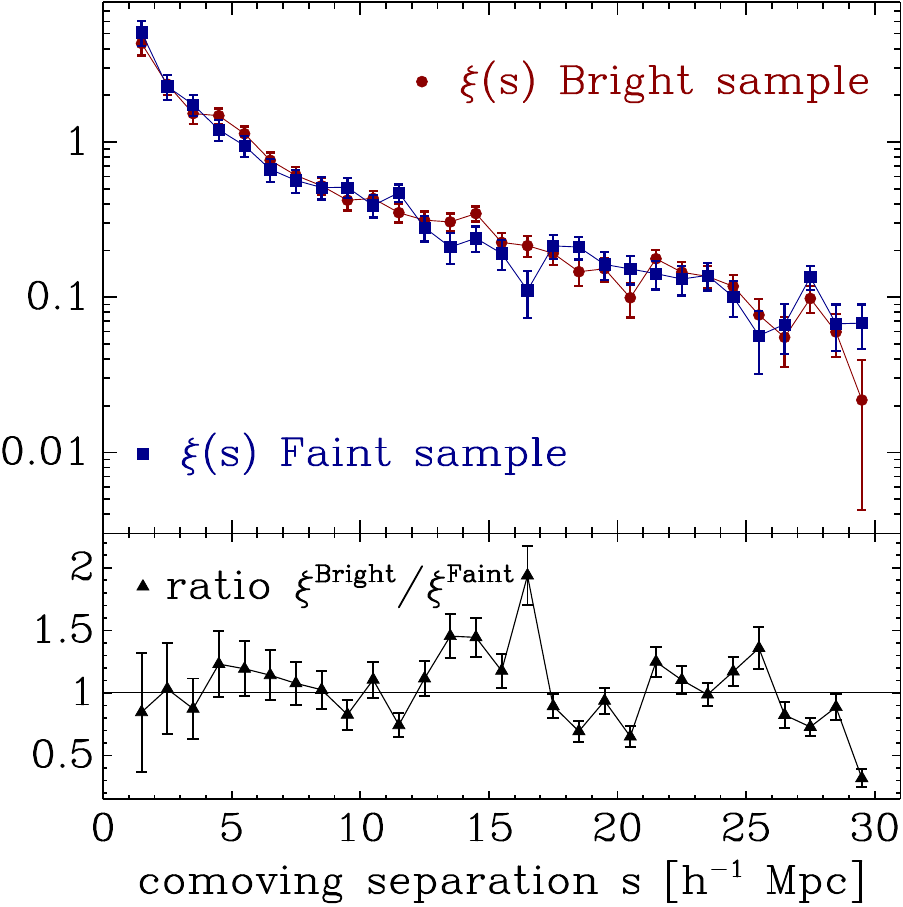}}
\caption{Comparison of the 3D two-point correlation function monopole $\xi(s)$ for the faint and bright samples. 
The top panel shows the measured correlations. The bottom panel displays the ratio between bright and faint samples.}
\label{fig:brightoverfaint}
\end{figure}

A comparison of the clustering amplitudes for the bright and faint samples is shown in Figure~\ref{fig:brightoverfaint}.
The error bars were computed propagating the uncertainties on the 2-point correlations.
We see a statistical preference for the clustering amplitude of the bright sample to be higher 
than the one of the faint sample. 
This is consistent with the power law parameters fitted, Table~\ref{tab:spatial_clustering}, the bright sample 
has higher values for clustering length $s_{0}$ than the faint sample, while the slope $\gamma$ seems more similar between samples. 
We do not draw any conclusions since we expect cosmic variance to have a large impact in these results.
%%%%%%%%%%%%%%%%%%%%%%%%%%%%%%%%%%%%%%%%%%%%%%%%%%%%%%%%%%%%%%
%%%%%%%%%%%%%%%%%%%%%%%%%%%%%%%%%%%%%%%%%%%%%%%%%%%%%%%%%%%%%%%%%%%%%%%%%%%%%%%%%%%%%%%%%%%%%%%%%%

\begin{figure}
\resizebox{\hsize}{!}{\includegraphics{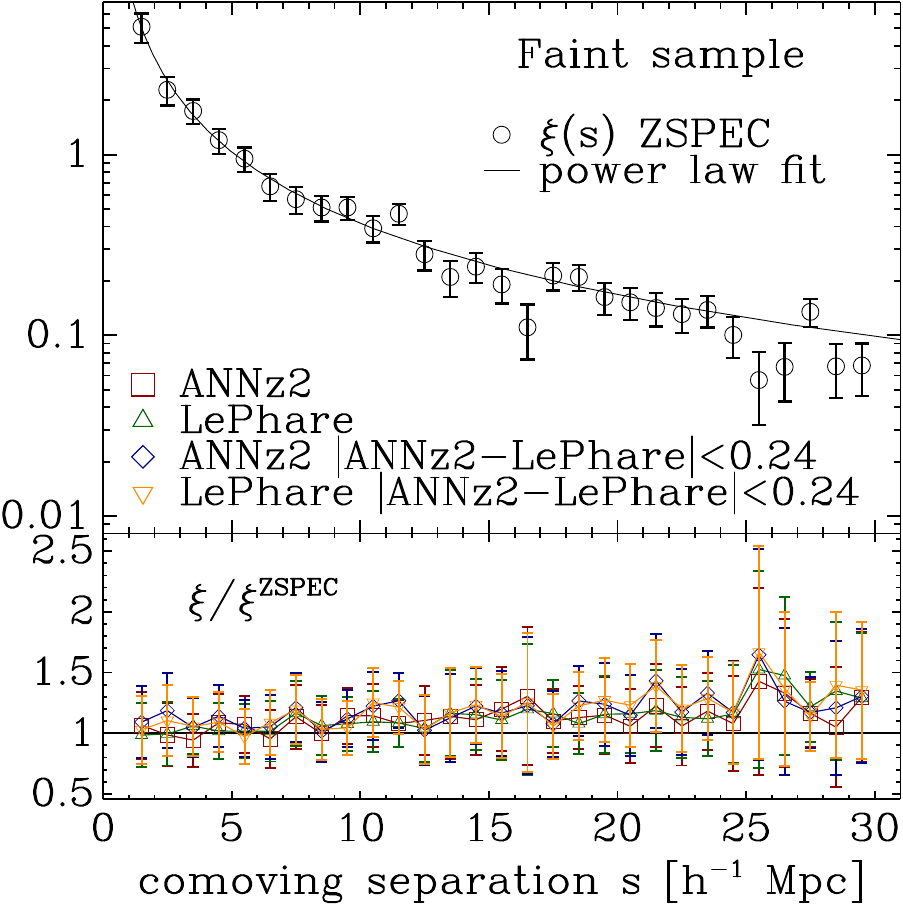}}
\caption{Comparison of $\xi(s)$ for different redshift selections for the faint sample only. The top panel shows the monopole for the 
clean $z_{sp}$ sample between $0.6<z<1.2$. The bottom panel shows the ratio between different redshift selections and the clean sample. 
An apparent increase on the clustering at large scales is seen for the photometric redshifts selections.}
\label{fig:faint_xiscomparison}
\end{figure}

Finally we find a slight increase on the clustering amplitude when photometric redshifts are considered. 
Figure~\ref{fig:faint_xiscomparison} compares the amplitudes of $\xi(s)$ when $z_{ph}$ are considered with respect to the spectroscopic selection.

%%%%%%%%%%%%%%%%%%%%%%%%%%%%%%%%%%%%%%%%%%%%
%%%%%%%%%%%%%%%%%%%%%%%%%%%%%%%%%%%%%%%%%%%%%%%%%%%%%%%%%%%%%%%%%%%%%%%%%%%%%%%%%%%%%%%%%%%%%%%%%%

%%%%%%%%%%%%%%%%%%%%%%%%%%%%%%%%%%%%%%%%%%%%%%%%%%%%%%%%%%%%%%%%%%

\subsection{Galaxy bias}
%%%%%%%%%%%%%%%%%%%%%%%%%%
\begin{figure*}
\includegraphics[width=0.8\textwidth,height=0.4\textwidth]{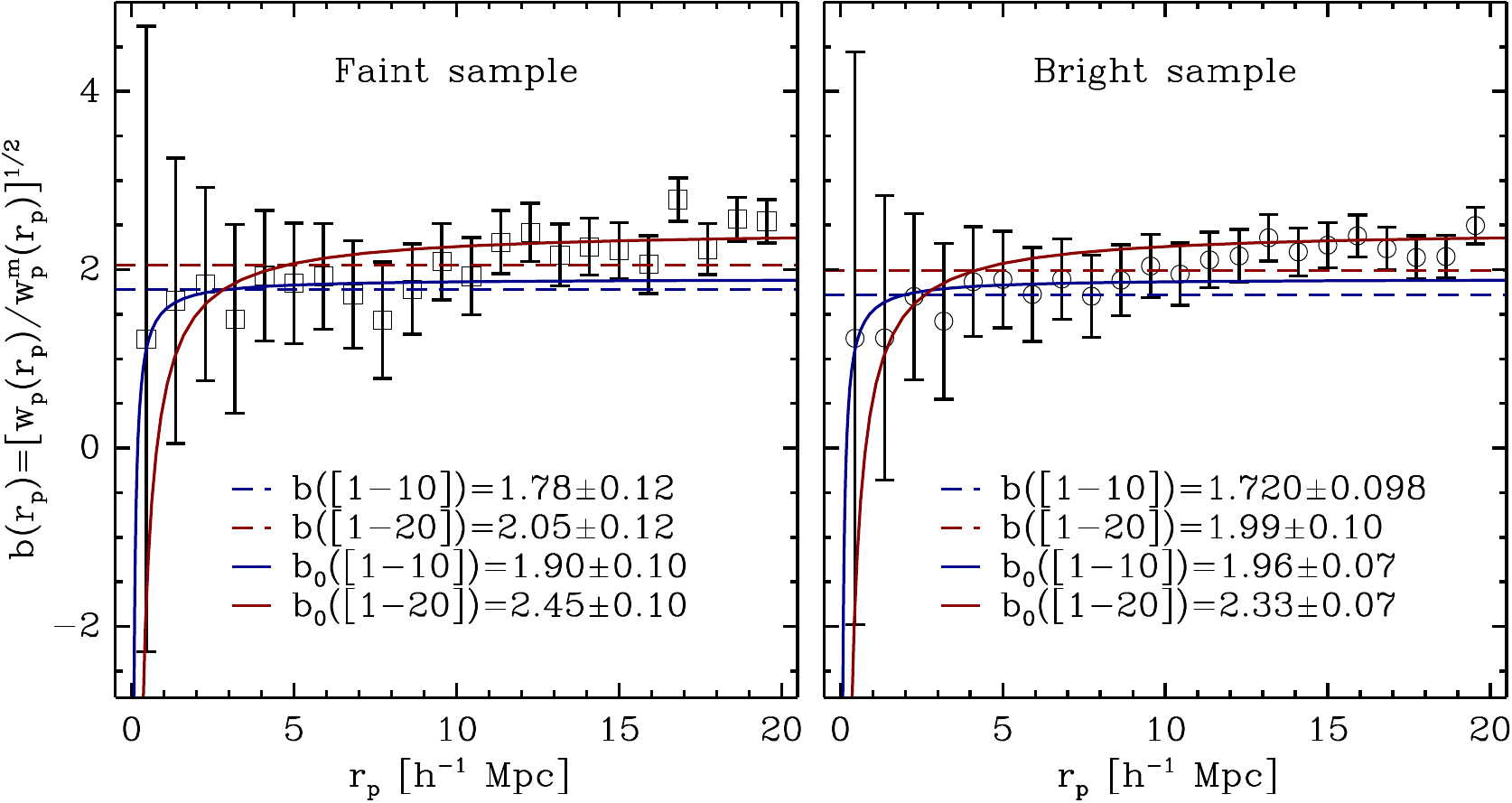}
\caption{The galaxy bias calculated using a constant and a scale dependent relation for the faint (left) and bright (right) sample,
measured from projected correlation function. In dashed lines we show the bias value calculated as the average between 1h-1Mpc
and 10h-1Mpc (blue) and between 1h-1Mpc and 20h-1Mpc (red). The straight lines are the scale dependence bias, fitting
to $b(r_{\rm p})=b_{0}+b_{1}/r_{\rm p}$, such that $b_{0}$ represents the large scale bias.
The results depend on the range of scales used in the average and in the limiting value of the fit. For comparison to previous
studies, we select the averaged bias between 1h-1Mpc and 10h-1Mpc (dashed blue) as our bias proxy, but we note that
different definitions gives different results.}
\label{fig:bias_r}
\end{figure*}
%%%%%%%%%%%%%%%%%%%%%%%%%%%%%%%%%%%%
In order to account for redshift space distortions, we follow the results from the VIPERS clustering analysis \citep{Marulli2013} 
and estimate the galaxy bias for our samples using the projected real space correlation function $w_{\rm p}(r_{\rm p})$. The galaxy bias is defined as
$$
b(r_{p})=\sqrt{\frac{w_{\rm p}(r_{p})}{w_{\rm p}^{\rm m}(r_{p})}}\,,
$$
where $w_{\rm p}(r_{p})$ is given by
$$
w_{\rm p}(r_{p}) = 2 \int^{\pi_{max}}_{0} d\pi' \xi(r_{p},\pi')\,,
$$
and is obtained from the galaxy sample, while $w_{\rm p}^{\rm m}(r_{p})$ is the projected correlation function of matter. 
We compute $w_{\rm p}^{\rm m}(r_{p})$ from the theoretical power spectrum obtained using CAMB \citep{Lewis:2002ah}, with the 
HALOFIT routine \citep{Smith:2002dz} for non-linear corrections assuming 
%for matter, $w^{\rm m}$, 
a flat $\Lambda$CDM model with $\Omega_{\rm m}=0.307$ and $\sigma_{8}=0.8$.

We measure the 3D correlation function $\xi(r_{p},\pi')$ in the spatial range $[1-40] \ h^{-1}$Mpc and integrate along the line of 
sight to obtain the measured $w_{\rm p}(r_{p})$ for all samples.

For comparison with VIPERS \citep{Marulli2013}, the bias is first estimated as the average of $b(r_{p})$ in the 
range of $[1-10]  h^{-1}$Mpc, where the bias is fairly constant, as claimed in VIPERS and shown in Figure~\ref{fig:bias_r}. 
In order to account for a small scale-dependency on the smallest scales, we also fit a relation $b(r_{p})=b_{0}+b_{1}/r_{p}$, 
such that $b_0$ can be taken as an estimate of the linear large-scale bias.

For both the averaged bias and the fit, the results depend on the scales used, as can be 
seen in Table~\ref{tab:wrp_resuls}. 
The smallest scales bring in nonlinearities whereas the largest scales are subject to sample variance, lower signal-to-noise and the 
largest possible effects from the $IC$. We note that we find the constant $IC$ ($0.01-0.02$) to be an order of magnitude lower than 
the correlation $(0.1-0.2$) around 20 $h^{-1}$Mpc. This 10\% effect on the correlation could in principle affect the bias estimation. This effect is 
smaller around 10 $h^{-1}$Mpc, where the correlation is a factor of 2-3 larger. 

The values for the bias change significantly between the average bias and the fit to the scale dependent bias, even when the same maximum scale 
is used in both procedures, indicating a measurable effect of the nonlinearity on the smallest scales. Finally, the bias values for the 
faint and bright samples agree within the error bars for all estimations. 
%For comparison with the VIPERS results, we use our averaged bias as reference. 
For the linear large-scale bias definition, the bright sample bias is higher than the faint, which is counterintuitive. 
This is explained by the limited area of our sample (small area and big cosmic variance) and indicates that we do not have 
the power yet to measure robustly any galaxy bias evolution between the two target selections. 
Galaxy bias estimations for the faint and bright sample are within 1$\sigma$. 
Nonetheless, we still can compare the broad behaviour of the galaxy bias against previous measurements.  
The errors have been obtained propagating the uncertainties in $r_{0}$ and $\gamma$ shown in the table, after fitting to \citep{Marulli2013}.
$$
w(r_{p})=r_{p}\left(\frac{r_{0}}{r_{p}}\right)^{\gamma} \frac{\Gamma(\frac{1}{2})\Gamma(\frac{\gamma-1}{2})}{\Gamma(\frac{\gamma}{2})}
$$

\begin{table*}
  \centering
  \begin{tabular}{ccccccc}
    \hline
    Sample & $s_{0}\; [h^{-1} {\rm Mpc}]$ & $\gamma$ & mean $z$ & bias averaged up to $[10-20] \ h^{-1}$Mpc & $\chi^{2}/$d.o.f & $b_{0}$ fitted up to $[10-20] \ Mpch^{-1}$\\
    \hline
    Bright & $4.18\pm0.26$ & $1.482\pm0.041$ & $0.855$ & $[1.72-1.99] \pm0.098$ & $2.80$ & $[1.96-2.33] \pm 0.07$ \\
    Faint  & $4.34\pm0.32$ & $1.501\pm0.049$ & $0.901$ & $[1.78-2.05] \pm0.12$   & $3.17$ & $[1.90-2.45] \pm 0.1$ \\

    \hline
  \end{tabular}
  \caption{Clustering properties and bias for the faint and bright samples selected with spectroscopic redshifts. 
The clustering length and slope were obtained by fitting a power law for $w(r_{p})$ for $0.5<r_{\rm p}<20 h^{-1}{\rm Mpc}$. 
The averaged bias value was obtained by averaging the scale dependent 
bias $b(r_{\rm p})=\left[ w(r_{\rm p})/w_{m}(r_{\rm p}) \right]^{1/2}$, while $b_{0}$ comes from a fit to the scale 
dependent bias $b(r_{\rm p})=b_{0}+b_{1}/r$. Both the average and the fit bias are obtained over scales $1<r_{\rm p}<10 h^{-1}{\rm Mpc}$ 
  as well as $1<r_{\rm p}<20 h^{-1}{\rm Mpc}$. }
  \label{tab:wrp_resuls}
\end{table*}

\begin{figure}\resizebox{\hsize}{!}{\includegraphics{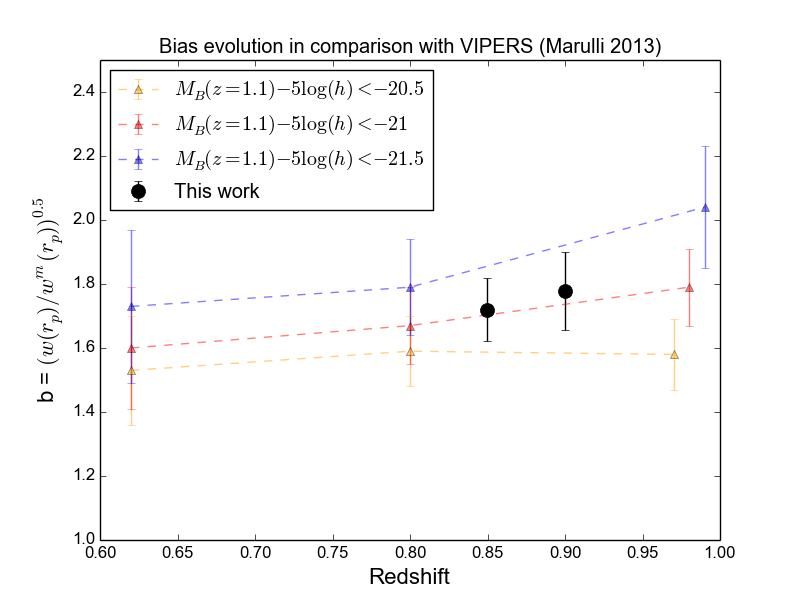}}
\caption{In black, the galaxy bias for our target selection samples in the range $0.6<z<1.2$ for the faint and bright. 
The reference values comes from the Table 1 of \citet{Marulli2013} from the VIPERS survey. 
In both cases biases have been measured as the average in $[1-10] \ h^{-1}$Mpc. 
Our bias agrees within one sigma with a galaxy population brighter than $M_{B}-5\log(h) < -21.0$.}
\label{fig:bb}
\end{figure}

In Figure~\ref{fig:bb}, we compare our measurements to those published for VIPERS \citep{Marulli2013}. 
For this comparison, we use our averaged bias as reference to reflect the VIPERS procedure.

\begin{figure}\resizebox{\hsize}{!}{\includegraphics{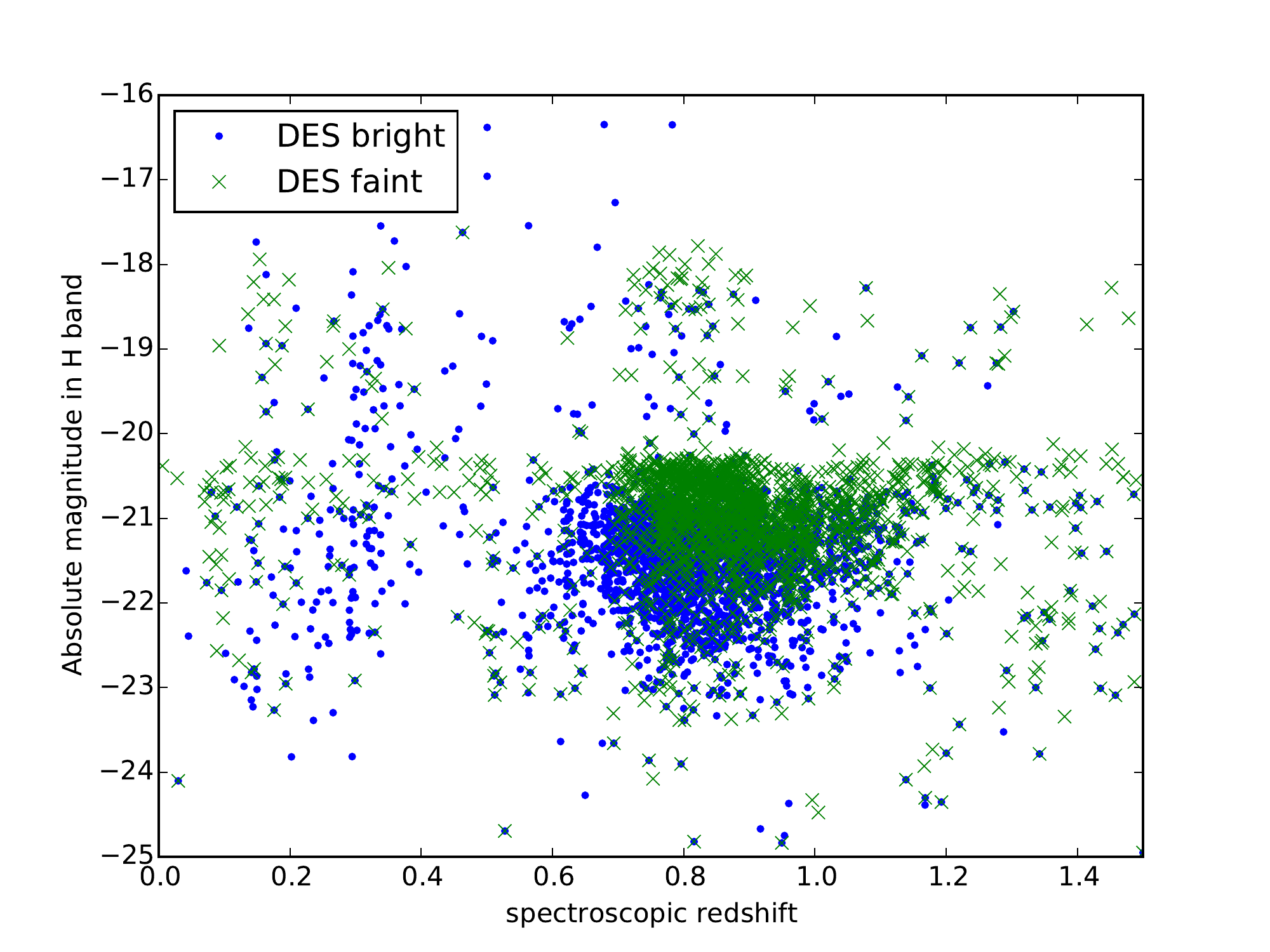}}
\caption{Absolute magnitude in the B-band for the ELG spectroscopic sample as a function of spectroscopic redshift, 
calculated using Lephare with the same configuration as in the photometric redshifts section. It agrees with a population $M_{B}-\log(h) < -21$.
in the redshift interval $0.6<z<1.2$, as it was predicted when we measured the galaxy bias.}
\label{fig:abs_mag}
\end{figure}

Our bias agrees with that from VIPERS for a population brighter than $M_{B}-\log(h) < -21$.
To confirm this result, we calculate the absolute magnitude for the faint and bright samples together 
(there is a strong overlap between both samples) to directly measure the limiting absolute magnitude of our sample.
We calculate the absolute magnitude for the B band using the template fiting code Lephare, fixing the redshift to its
spectroscopic value. We show the B-band absolute magnitude density distribution in figure~\ref{fig:abs_mag} as a function of redshift for the 
bright and faint sample. The result agrees well with what it is expected from the galaxy bias of the sample. 
This result should be taken with caution as we have not included the cosmic variance uncertainty. 
The luminosity dependent clustering will be analysed in more detail in future studies.
%%%%%%%%%%%%%%%%%%%%%%%%%%%%%%%%%%%%%%%%%%%%%%%%%%%%%%%%%%%%%%%%%%%%%%%%%%%%%%%%%%%%%%%%%%%%%%%%%%
%%%%%%%%%%%%%%%%%%%%%%%%%%%%%%%%%%%%%%%%%%%%%%%%%%%%%%%%%%%%%%%%%%%%%%%%%%%%%%%%%%%%%%%%%%%%%%%%%%
%%%%%%%%%%%%%%%%%%%%%%%%%%%%%%%%%%%%%%%%%%%%%%%%%%%%%%%%%%%%%%%%%%%%%%%%%%%%%%%%%%%%%%%%%%%%%%%%%%
%%%%%%%%%%%%%%%%%%%%%%%%%%%%%%%%%%%%%%%%%%%%%%%%%%%%%%%%%%%%%%%%%%%%%%%%%%%%%%%%%%%%%%%%%%%%%%%%%%
\section{Conclusion}
We have used 9.2 deg$^2$ of eBOSS test plates data to study the properties of different possible ELG target selections. 
We design four target selections using SDSS, SCUSS, and DES-SVA1 data. 
We study the SDSS-SCUSS selections in \citep{Delubac15,Comparat15} in preparation. 
The bright DES grz bands selection achieves 73\% success rate and 71\% in the desired redshift
window $0.6<z<1.2$. The faint DES grz bands selection have sligthly lower performances with 66\% success rate and 68\% in the
redshift window. Both selections have a stellar contamination lower than 2\%. We find a mean redshift of 0.8 and 0.87
for respectively the bright and faint selection.\\ 
To prepare the eBOSS survey, we looked at the strongest systematic effects which can affect
the power spectrum measurement: stellar photometry contamination, airmass, galactic dust, and survey depth
 accross the DES year one data. We find a galaxy density variation lower than 15\% for each of these systematic effects,
which is the highest fluctuation allowed to avoid damaging measurements, as studied in \citet{Dawson15}. 
With a target density of 857 galaxy/deg$^2$, our analysis suggests the DES bright selection will give the 
most accurate power spectrum measurement with an eBOSS-like survey type.\\ 
With the 4600 eBOSS $z_{sp}$, we investigate possible techniques to identify the $z_{ph}$ outliers. The 
outlier fraction is one of the biggest source of systematics in cosmic shear and large scale structure analysis
\citep{Bernstein09}. With a five optical broadband photometric survey, we need to identify and control
the outlier fraction. Locating color boxes with a high percentage of outliers is a possible way to deal with this.
Using the random forest code TPZ, we find that removing the color branches with a a percentage of outliers higher than 10\%,
we are left with 10\% outliers and a galaxy sample of 71\% completness. 
\citet{Newman13b} suggests another possible technique to decrease the outlier fraction using a comparison between
template fitting $z_{ph}$ and machine learning $z_{ph}$. We find that in reducing the galaxy sample by 15\% we decrease the
outlier fraction by 30\%.  

We investigated the clustering properties of our samples, estimating the 3D two-point correlation function monopole $\xi(s)$ 
and the projected real space correlation function $w(r_{p})$. We used these measurements to compute the large-scale galaxy bias, 
and found it to be consistent with previous ELG measurements. The galaxy bias between the DES bright and faint sample are 
within 1$\sigma$ of each other. 
We find a slightly higher bias for the faint sample compared to the bright which is expected due to redshift evolution. 
We also looked at the binning effect in clustering analysis when having to define a redshift window with photometric redshifts. 
Considering that DES will have a good photomeric redshifts calibration, we used spectroscopic redshifts to
compute correlation functions and use the photo-z to define the 0.6 to 1.2 redshift window. We
do not find significant differences when using spectroscopic and photometric redshifts.
We finally compare the mean value of the galaxy bias to the deep spectroscopic survey VIPERS and find a good agreement. 
 
\bibliographystyle{mn2e}
\bibliography{biblio_mnras}
\section{Acknowledgements}
HC is supported by CNPq. 
AC thanks Fernando de Simoni for useful discussions. 
AC acknowledges financial support provided by the PAPDRJ CAPES/FAPERJ Fellowship. 
ML is partially supported by FAPESP and CNPq. 
FBA acknowleges the support of the Royal society via a RS University Research Fellowhsip.
FS acknowledges financial support provided by CAPES under contract No. 3171-13-2.
JC, and FP acknowledge support from the Spanish MICINN grant MultiDark CSD2009-00064, MINECO Severo Ochoa Programme grant SEV-2012-0249, and grant AYA2014-60641-C2- 1-P.  FP wish to thank the Lawrence Berkeley National Laboratory for the hospitality and the Spanish MEC “Salvador de Madariaga” program, Ref. PRX14/00444.

This paper has gone through internal review by the DES collaboration.
We are grateful for the extraordinary contributions of our CTIO colleagues and the DECam Construction, Commissioning and Science Verification
teams in achieving the excellent instrument and telescope conditions that have made this work possible. The success of this project also 
relies critically on the expertise and dedication of the DES Data Management group.
Funding for the DES Projects has been provided by the U.S. Department of Energy, the U.S. National Science Foundation, the Ministry of Science and Education of Spain, the Science and Technology Facilities Council of the United Kingdom, the Higher Education Funding Council for England, the National Center for Supercomputing Applications at the University of Illinois at Urbana-Champaign, the Kavli Institute of Cosmological Physics at the University of Chicago, the Center for Cosmology and Astro-Particle Physics at the Ohio State University, the Mitchell Institute for Fundamental Physics and Astronomy at Texas A\&M University, Financiadora de Estudos e Projetos, Fundaca\ ̃o Carlos Chagas Filho de Amparo \`a Pesquisa do Estado do Rio de Janeiro, Conselho Nacional de Desenvolvimento Cient\'ifico e Tecnologico and the Minis\'erio da Ciencia e Tecnologia, the Deutsche Forschungsgemeinschaft and the Collaborating Institutions in the Dark Energy Survey. \\

The DES data management system is supported by the National Science Foundation under Grant Number AST1138766. The DES participants from Spanish institutions are partially supported by MINECO under grants AYA201239559, ESP2013-48274, FPA2013-47986, and Centro de Excelencia Severo Ochoa SEV-2012-0234, some of which include ERDF funds from the European Union. 

The Collaborating Institutions are Argonne National Laboratory, the University of California at Santa Cruz, the University of Cambridge, Centro de Investigaciones Energeticas, Medioambientales y Tecnologicas-Madrid, the University of Chicago, University College London, the DES-Brazil Consortium, the Eidgenössische Technische Hochschule (ETH) Zürich, Fermi National Accelerator Laboratory, the University of Edinburgh, the University of Illinois at Urbana-Champaign, the Institut de Ci\`encies de l’Espai (IEEC/CSIC), the Institut de F\'isica d’Altes Energies, Lawrence Berkeley National Laboratory, the Ludwig-Maximilians Universität and the associated Excellence Cluster Universe, the University of Michigan, the National Optical Astronomy Observatory, the University of Nottingham, The Ohio State University, the University of Pennsylvania, the University of Portsmouth, SLAC National Accelerator Laboratory, Stanford University, the University of Sussex, and Texas A\&M University.

Funding for the Sloan Digital Sky Survey IV has been provided by
the Alfred P. Sloan Foundation, the U.S. Department of Energy Office of
Science, and the Participating Institutions. SDSS-IV acknowledges
support and resources from the Center for High-Performance Computing at
the University of Utah. The SDSS web site is www.sdss.org.

SDSS-IV is managed by the Astrophysical Research Consortium for the 
Participating Institutions of the SDSS Collaboration including the 
Brazilian Participation Group, the Carnegie Institution for Science, 
Carnegie Mellon University, the Chilean Participation Group, the French Participation Group, Harvard-Smithsonian Center for Astrophysics, 
Instituto de Astrof\'isica de Canarias, The Johns Hopkins University, 
Kavli Institute for the Physics and Mathematics of the Universe (IPMU) / 
University of Tokyo, Lawrence Berkeley National Laboratory, 
Leibniz Institut f\"ur Astrophysik Potsdam (AIP),  
Max-Planck-Institut f\"ur Astronomie (MPIA Heidelberg), 
Max-Planck-Institut f\"ur Astrophysik (MPA Garching), 
Max-Planck-Institut f\"ur Extraterrestrische Physik (MPE), 
National Astronomical Observatory of China, New Mexico State University, 
New York University, University of Notre Dame, 
Observat\'ario Nacional / MCTI, The Ohio State University, 
Pennsylvania State University, Shanghai Astronomical Observatory, 
United Kingdom Participation Group,
Universidad Nacional Aut\'onoma de M\'exico, University of Arizona, 
University of Colorado Boulder, University of Oxford, University of Portsmouth, 
University of Utah, University of Virginia, University of Washington, University of Wisconsin, 
Vanderbilt University, and Yale University.

\appendix
\section{eBOSS systematics}
%%%%%%%%%%%%%%%%%%%%%%%%%%%%%%%%%%%%%%%%%%%%%%%%%%%%%%%%%%%%%%%%%%%%%%%%%%%%%%%%%%%%%%%%%%%%%%%%%%
\begin{figure*}
\hbox{
\includegraphics[width=0.35\textwidth,height=0.35\textwidth]{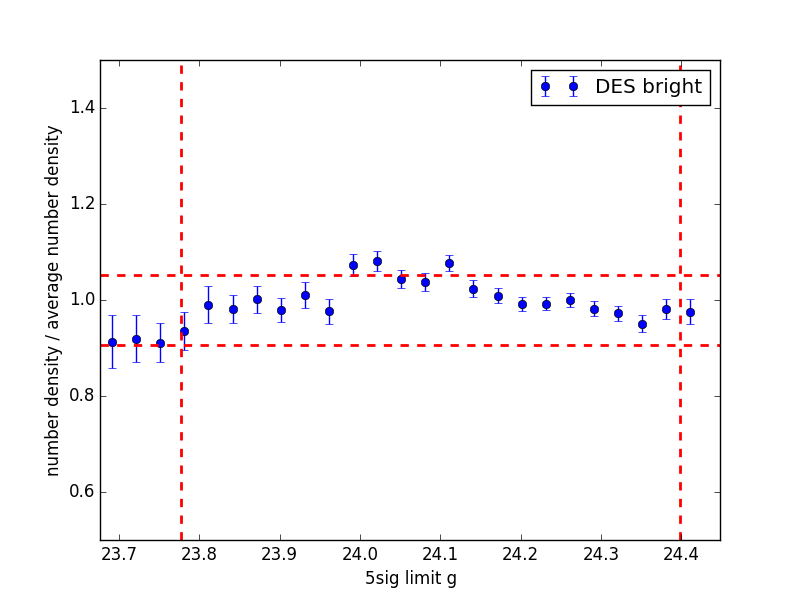}
\includegraphics[width=0.35\textwidth,height=0.35\textwidth]{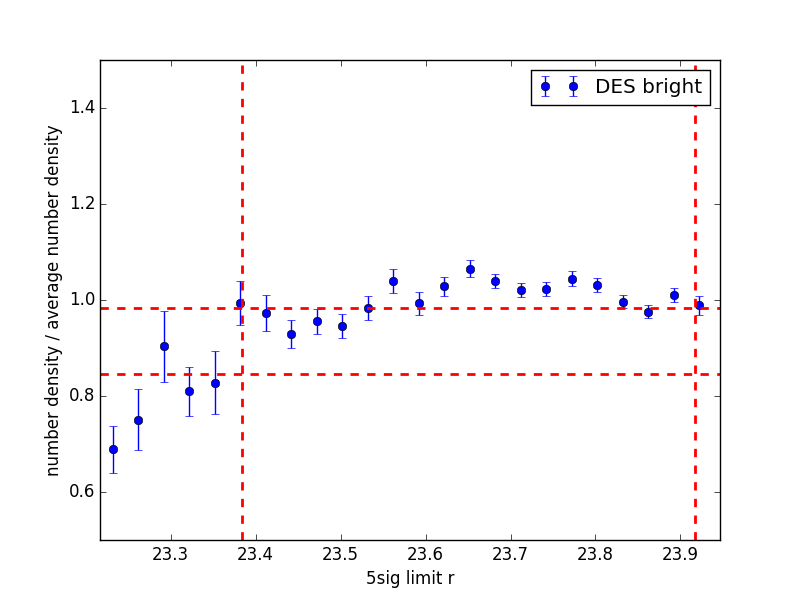}
\includegraphics[width=0.35\textwidth,height=0.35\textwidth]{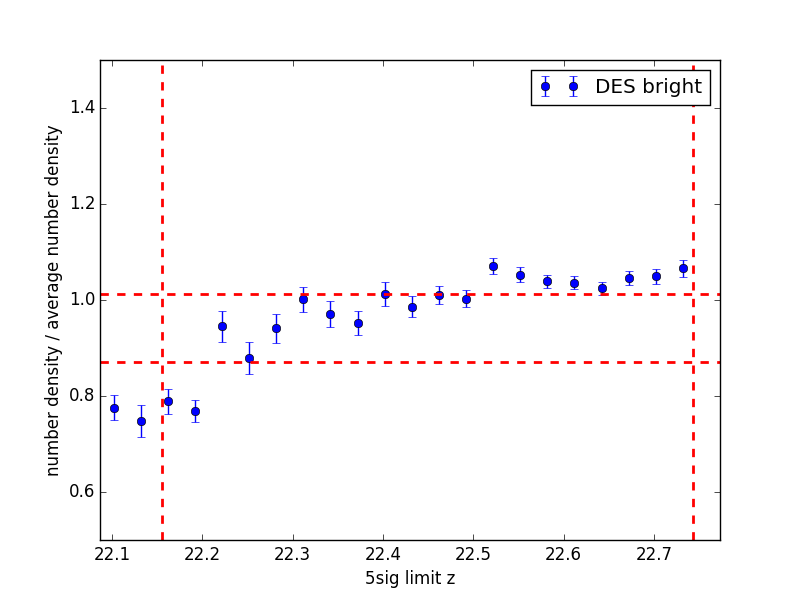}
}
\hbox{
\includegraphics[width=0.35\textwidth,height=0.35\textwidth]{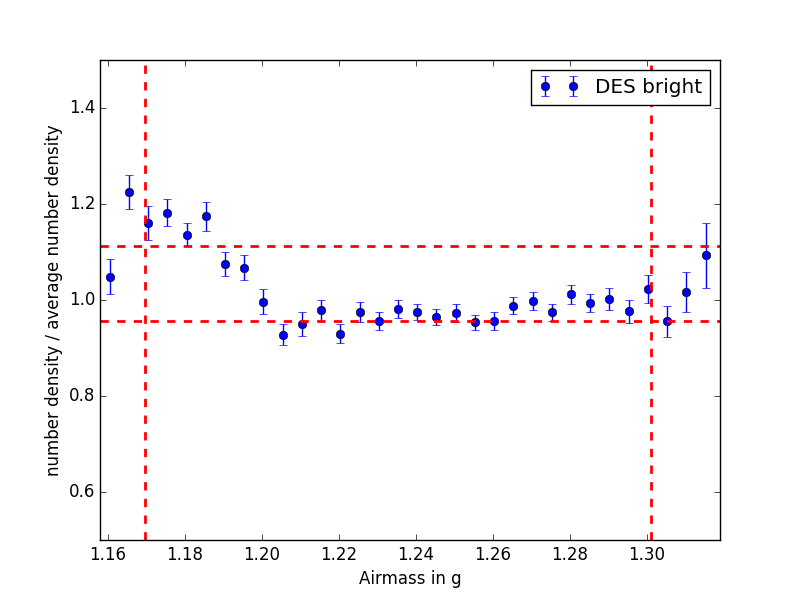}
\includegraphics[width=0.35\textwidth,height=0.35\textwidth]{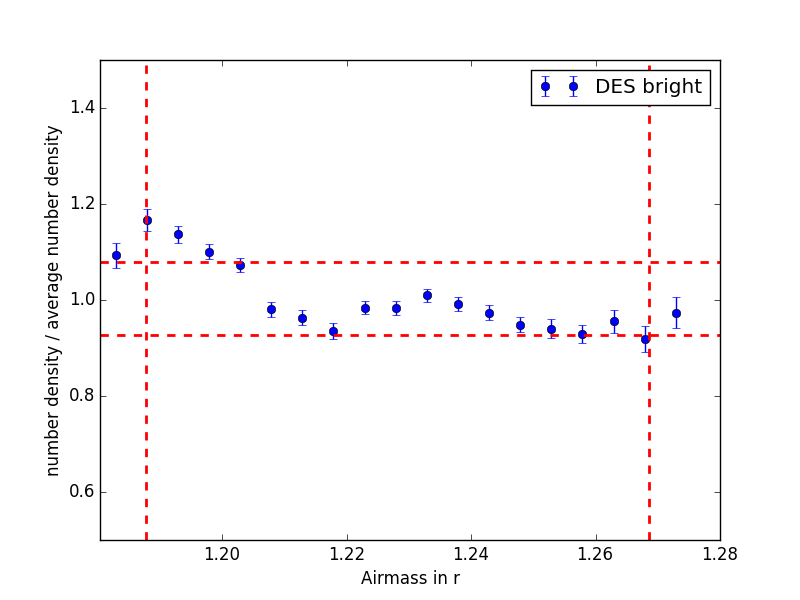}
\includegraphics[width=0.35\textwidth,height=0.35\textwidth]{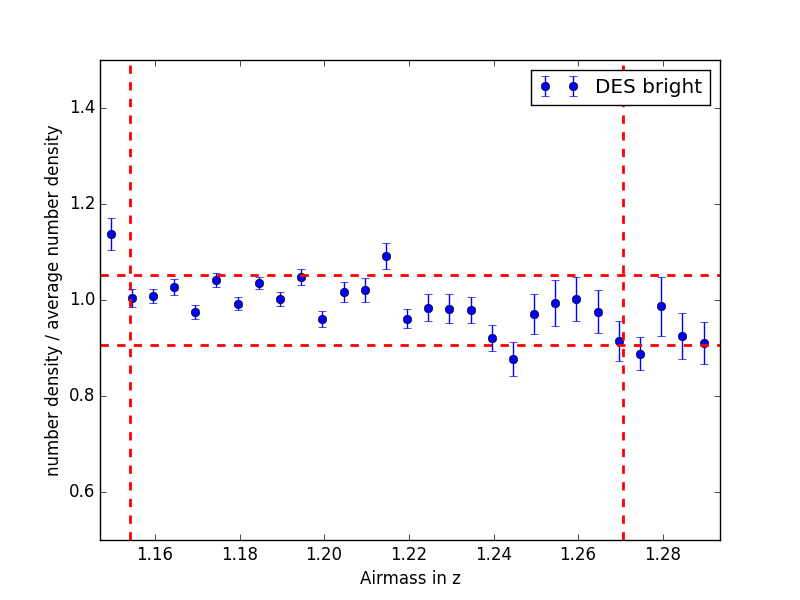}
}
\hbox{
\includegraphics[width=0.35\textwidth,height=0.35\textwidth]{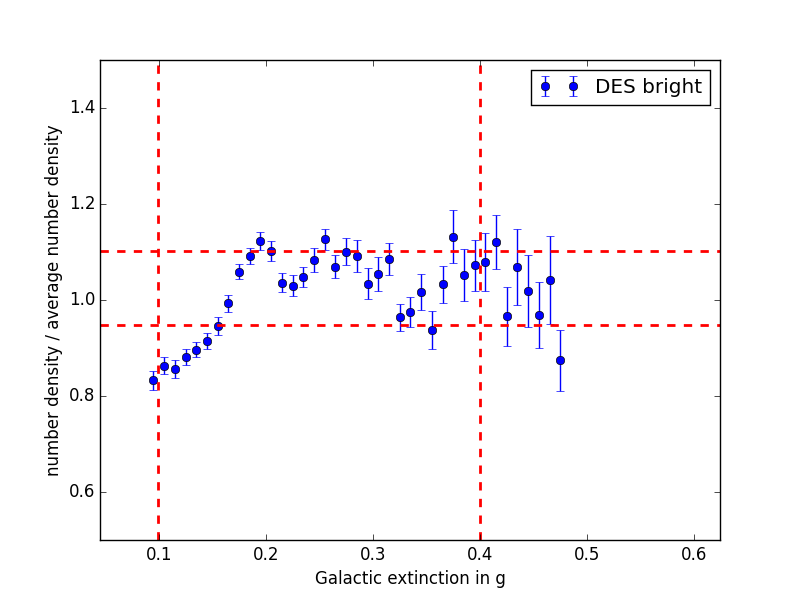}
\includegraphics[width=0.35\textwidth,height=0.35\textwidth]{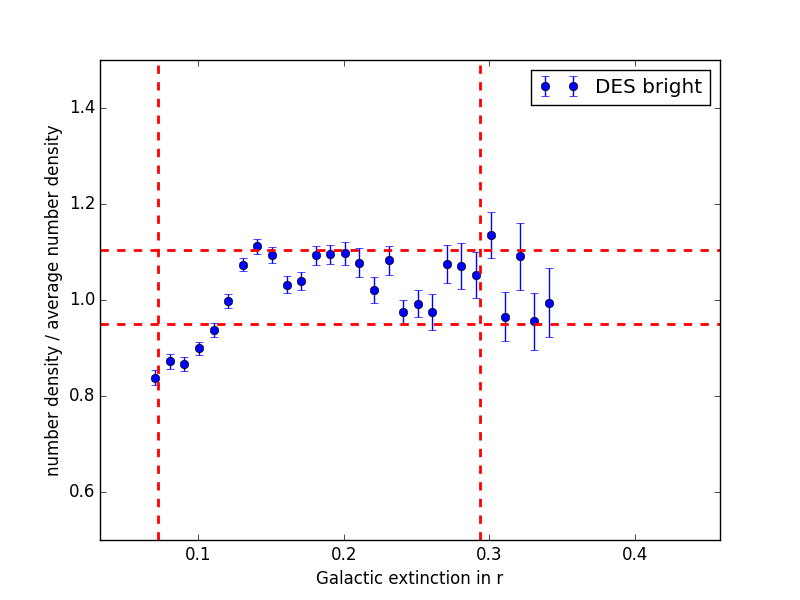}
\includegraphics[width=0.35\textwidth,height=0.35\textwidth]{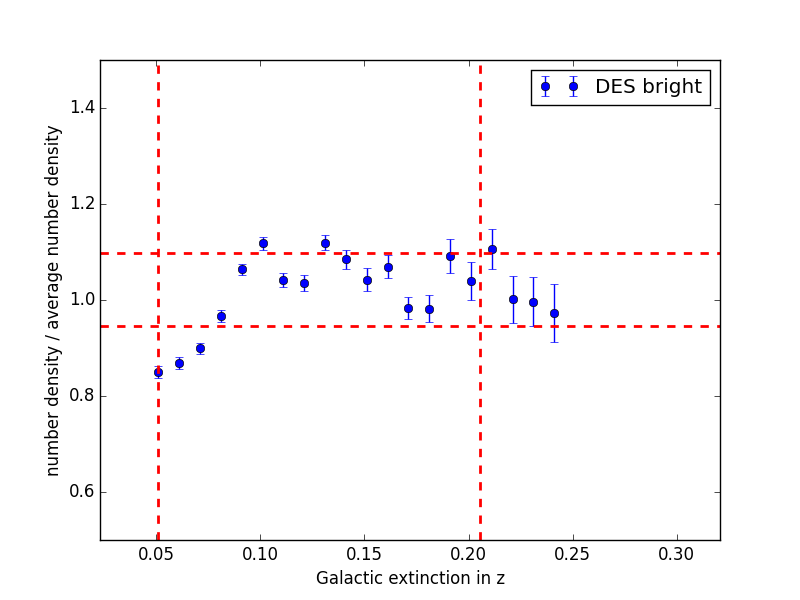}
}
\caption{Density fluctuation of galaxies as a function of the depth, airmass and Galactic extinction in the top, middle and bottom row for the g,r,z bands. The two vertical and horizontal red axes show respectively the 5 and 95\% of the depth, airmass, galactic extinction distribution and 15\% around the mean galaxy density fluctuation for the eBOSS target selection.}
\label{fig:sys_eboss}
\end{figure*}
%%%%%%%%%%%%%%%%%%%%%%%%%%%%%%%%%%%%%%%%%%%%%%%%%%%%%%%%%%%%%%%%%%%%%%%%%%%%%%%%%%%%%%%%%%%%%%%%%%
\newpage
\section*{Affiliations}
{\small
$^{1}$ Department of Physics and Astronomy, University College London, 132 Hampstead road, London NW1 2PS, UK \\
$^{2}$ Laboratoire d’Astrophysique, Ecole polytechnique Federale de Lausanne, 1015 Lausanne, Switzerland\\
$^{3}$ Instituto de F\'isica Teo\'rica, (UAM/CSIC), Universidad Aut\'onoma de Madrid, Cantoblanco, E-28049 Madrid, Spain \\
$^{4}$ Departamento de F\'isica Te\'orica M8, Universidad Aut\'onoma de Madrid (UAM), Cantoblanco, E-28049, Madrid, Spain \\
$^{5}$ Departamento de F\'isica Matem\'atica, Instituto de F\'isica, Universidade de S\~ao Paulo, S\~ao Paulo-SP, Brazil \\
$^{6}$ Laborat\'orio Interinstitucional de e-Astronomia-LIneA, Rua Gal. Jos\'e Cristino 77, Rio de Janeiro, RJ 20921-400, Brazil \\
$^{7}$ Observat\'orio Nacional, Rua Gal. Jos\'e Cristino 77, Rio de Janeiro, RJ 20921-400, Brazil \\
$^{8}$ Fermi National Accelerator Laboratory, P. O. Box 500, Batavia, IL 60510, USA \\
$^{9}$ Department of Physics and Electronics, Rhodes University, PO Box 94, Grahamstown 6140 South Africa \\
$^{10}$ Campus of International Excellence UAM+CSIC, Cantoblanco, E-28049 Madrid, Spain \\
$^{11}$ Lawrence Berkeley National Laboratory, 1 Cyclotron Road, Berkeley, CA, 94720, USA \\
$^{12}$ Instituto de Astrof\'isica de Andaluc\'ia (CSIC), Glorieta de la Astronom\'ia, E-18080 Granada, Spain \\
$^{13}$ Department of Physics \& Astronomy, Johns Hopkins University, 3400 N. Charles Street, Baltimore, MD 21218, USA \\
$^{14}$ Hubble Fellow \\
$^{15}$ Instituto de Astronomıa, Universidad Nacional Autonoma de Mexico, A.P. 70-264, 04510, Mexico, D.F., Mexico\\
$^{16}$ Cerro Tololo Inter-American Observatory, National Optical Astronomy Observatory, Casilla 603, La Serena, Chile\\
$^{17}$ Institute of Astronomy, University of Cambridge, Madingley Road, Cambridge CB3 0HA, UK\\
$^{18}$ Kavli Institute for Cosmology, University of Cambridge, Madingley Road, Cambridge CB3 0HA, UK\\
$^{19}$ CNRS, UMR 7095, Institut d'Astrophysique de Paris, F-75014, Paris, France\\
$^{20}$ Sorbonne Universit\'es, UPMC Univ Paris 06, UMR 7095, Institut d'Astrophysique de Paris, F-75014, Paris, France\\
$^{21}$ Institute of Cosmology \& Gravitation, University of Portsmouth, Portsmouth, PO1 3FX, UK\\
$^{22}$ Department of Astronomy, University of Illinois, 1002 W. Green Street, Urbana, IL 61801, USA\\
$^{23}$ National Center for Supercomputing Applications, 1205 West Clark St., Urbana, IL 61801, USA\\
$^{24}$ Institut de Ci\`encies de l'Espai, IEEC-CSIC, Campus UAB, Carrer de Can Magrans, s/n,  08193 Bellaterra, Barcelona, Spain\\
$^{25}$ Institut de F\'{\i}sica d'Altes Energies, Universitat Aut\`onoma de Barcelona, E-08193 Bellaterra, Barcelona, Spain\\
$^{26}$ Kavli Institute for Particle Astrophysics \& Cosmology, P. O. Box 2450, Stanford University, Stanford, CA 94305, USA\\
$^{27}$ Excellence Cluster Universe, Boltzmannstr.\ 2, 85748 Garching, Germany\\
$^{28}$ Faculty of Physics, Ludwig-Maximilians University, Scheinerstr. 1, 81679 Munich, Germany\\
$^{29}$ Department of Physics and Astronomy, University of Pennsylvania, Philadelphia, PA 19104, USA\\
$^{30}$ Jet Propulsion Laboratory, California Institute of Technology, 4800 Oak Grove Dr., Pasadena, CA 91109, USA\\
$^{31}$ Kavli Institute for Cosmological Physics, University of Chicago, Chicago, IL 60637, USA\\
$^{32}$ Department of Physics, University of Michigan, Ann Arbor, MI 48109, USA\\
$^{33}$ Max Planck Institute for Extraterrestrial Physics, Giessenbachstrasse, 85748 Garching, Germany\\
$^{34}$ Universit\""ats-Sternwarte, Fakult\""at f\""ur Physik, Ludwig-Maximilians Universit\""at M\""unchen, Scheinerstr. 1, 81679 M\""unchen, Germany\\
$^{35}$ Center for Cosmology and Astro-Particle Physics, The Ohio State University, Columbus, OH 43210, USA\\
$^{36}$ Department of Physics, The Ohio State University, Columbus, OH 43210, USA\\
$^{37}$ Australian Astronomical Observatory, North Ryde, NSW 2113, Australia\\
$^{38}$ George P. and Cynthia Woods Mitchell Institute for Fundamental Physics and Astronomy, and Department of Physics and Astronomy, Texas A\&M University, College Station, TX 77843,  USA\\
$^{39}$ Observat\'orio Nacional, Rua Gal. Jos\'e Cristino 77, Rio de Janeiro, RJ - 20921-400, Brazil\\
$^{40}$ Instituci\'o Catalana de Recerca i Estudis Avan\c{c}ats, E-08010 Barcelona, Spain\\
$^{41}$ SLAC National Accelerator Laboratory, Menlo Park, CA 94025, USA\\
$^{42}$ Department of Physics and Astronomy, Pevensey Building, University of Sussex, Brighton, BN1 9QH, UK\\
$^{43}$ Centro de Investigaciones Energ\'eticas, Medioambientales y Tecnol\'ogicas (CIEMAT), Madrid, Spain\\
$^{44}$ Instituto de F\'\i sica, UFRGS, Caixa Postal 15051, Porto Alegre, RS - 91501-970, Brazil\\
$^{45}$ Department of Physics, The Ohio State University, Columbus, OH 43210, USA\\
$^{46}$ Department of Physics, University of Illinois, 1110 W. Green St., Urbana, IL 61801, USA\\
$^{47}$ Department of Physics and Astronomy, University of Utah, 115 S 1400 E, Salt Lake City, UT 84112, USA\\
}
\end{document}